%% file: NeuroPath.tex
\documentclass[9pt,sigconf]{acmart}

\usepackage{fancyhdr}
\usepackage{enumitem}
\usepackage{stfloats}
\usepackage{url}
\usepackage{xspace}
\usepackage{xcolor}
\usepackage{soul}
\usepackage{pifont}
\usepackage{graphicx}
\usepackage{subfig}
\usepackage{array}

\usepackage[T1]{fontenc}
\usepackage{booktabs}   
\usepackage{multirow}   
\usepackage{siunitx}    
\usepackage{caption}    
\usepackage{tabularx}   

\newcolumntype{C}{>{\centering\arraybackslash}X}

\usepackage{listings}
\newcommand{\ie}{\emph{i.e.},\xspace}
\newcommand{\eg}{\emph{e.g.},\xspace}
\newcommand{\etc}{\emph{etc.}\xspace}

\graphicspath{{./}{Figure/}}

\newcommand{\parahead}[1]{\vspace*{1ex plus 0.25ex minus 0.25ex}\noindent {\bfseries #1}}

\newcommand{\systemname}{NeuroPath\xspace}

\begin{document}

\title{NeuroPath: Practically Adopting Motor Imagery Decoding through EEG Signals}

\author{Jiani Cao}
\affiliation{
  \institution{Department of Computer Science \\City University of Hong Kong}
  \city{Hong Kong}
  \country{China}
}
\email{jncao2-c@my.cityu.edu.hk}

\author{Kun Wang}
\affiliation{
  \institution{Department of Computer Science \\City University of Hong Kong}
  \city{Hong Kong}
  \country{China}
}
\email{kwang69-c@my.cityu.edu.hk}

\author{Yang Liu}
\authornote{Corresponding authors.}
\affiliation{
  \institution{Department of Computer Science \\Florida State University}
  \city{Tallahassee}
  \country{United States}
}
\email{yl25r@fsu.edu}

\author{Zhenjiang Li}
\authornotemark[1]
\affiliation{
  \institution{Department of Computer Science \\City University of Hong Kong}
  \city{Hong Kong}
  \country{China}
}
\email{zhenjiang.li@cityu.edu.hk}

\begin{abstract}
Motor Imagery (MI) is an emerging Brain–Computer Interface (BCI) paradigm in which a person imagines a body movement without any physical action. By decoding the scalp-recorded electroencephalography (EEG) signals, BCIs can establish direct communication pathways to control external devices, offering significant potential in prosthetics, rehabilitation, and human–computer interaction. However, existing solutions remain difficult to deploy in practice. (i) Most employ independent, opaque models for each MI task. This fragmented methodology lacks a unified architectural foundation. Consequently, these models are trained in isolation and fail to learn robust representations from diverse datasets, which often results in modest performance. (ii) They primarily adopt fixed sensor deployment, whereas real-world setups vary in electrode number and placement, causing models trained on one configuration to fail under another. (iii) Performance degrades sharply under low-SNR conditions typical of consumer-grade EEG. Together, these limitations hinder the practical adoption of MI-based BCIs.

To address these challenges, we present NeuroPath, a well-designed neural architecture for robust MI decoding. NeuroPath takes high-level inspiration from the brain's signal pathway from cortex to scalp, utilizing a deep neural architecture with specialized modules for signal filtering, spatial representation learning, and feature classification, enabling unified decoding rather than task-specific black-box learning. To handle variations in electrode configurations, we introduce a spatially aware graph adapter that accommodates different electrode numbers and placements. To enhance robustness under low-SNR conditions, NeuroPath incorporates multimodal auxiliary training to refine EEG representations and stabilize performance on noisy, real-world data. Evaluations on three self-collected consumer-grade datasets and three public medical-grade datasets demonstrate that NeuroPath achieves superior performance.
\end{abstract}


\acmYear{2026}\copyrightyear{2026}
\setcopyright{cc}
\setcctype[4.0]{by}
\acmConference[SenSys '26]{ACM/IEEE International Conference on Embedded Artificial Intelligence and Sensing Systems}{May 11--14, 2026}{Saint Malo, France}
\acmBooktitle{ACM/IEEE International Conference on Embedded Artificial Intelligence and Sensing Systems (SenSys '26), May 11--14, 2026, Saint Malo, France}
\acmDOI{10.1145/3774906.3802770}
\acmISBN{979-8-4007-2309-4/26/05}


\maketitle

\input{Body/intro}
\input{Body/pre}
\input{Body/design}

\input{Body/evaluation}

\input{Body/related}
\input{Body/conclusion}

\clearpage
\bibliographystyle{ACM-Reference-Format}
\bibliography{reference}

\end{document}

%% file: Body/intro.tex
\section{Introduction}
\label{sec:intro}

\begin{figure}[t]
	\centering
	\includegraphics[width=2.6in]{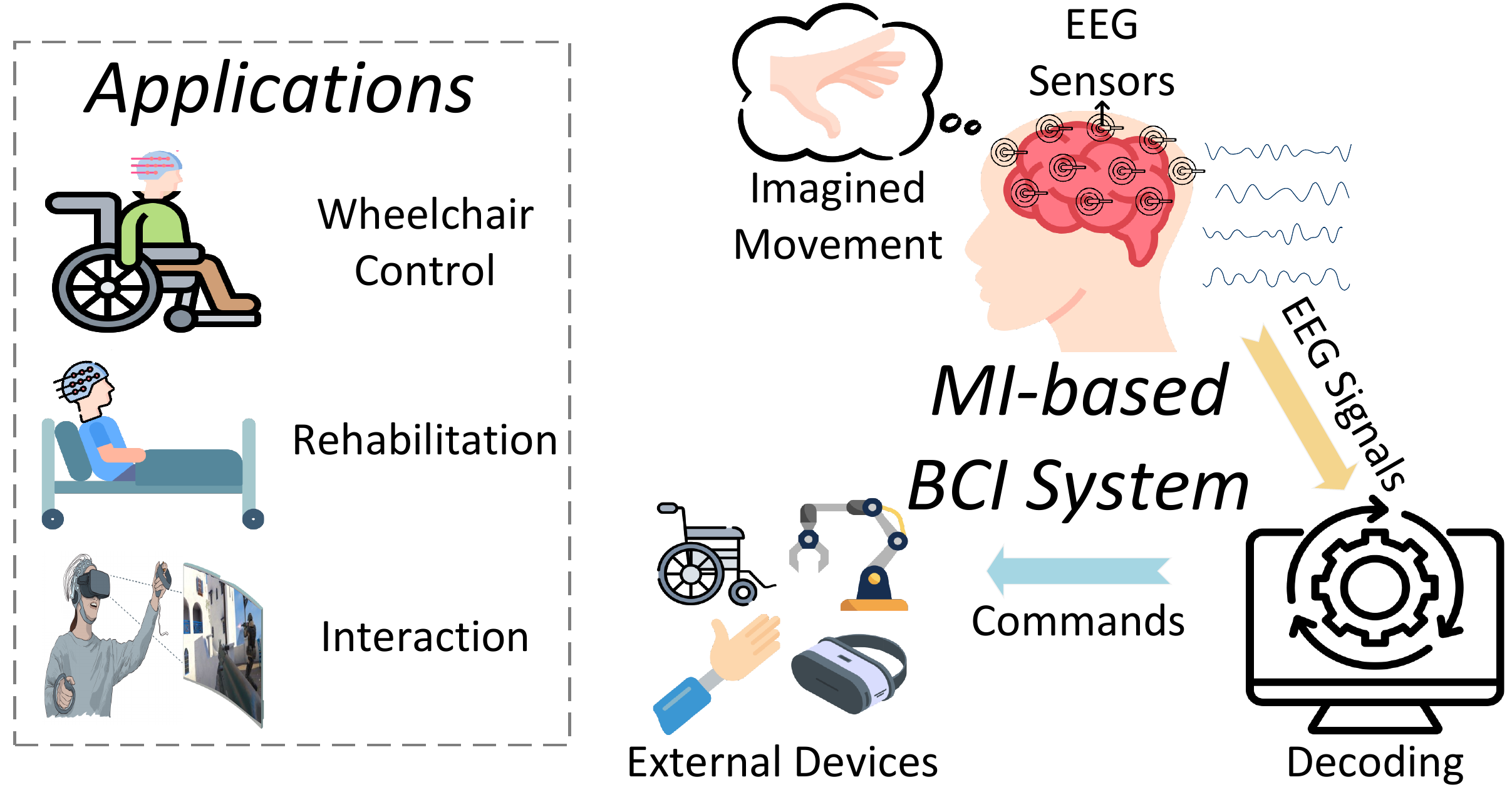}
    \caption{The process of MI-based BCI system and its potential applications.}
	\label{fig:scenarios}
	\vspace{-.2in}
\end{figure}

Brain–Computer Interface (BCI) technology~\cite{edelman2024non} enables direct communication between the brain and external devices, unlocking applications in assistive control, rehabilitation, and human–computer interaction~\cite{varbu2022past,lotte2012combining,islam2023recent}. Among BCI paradigms, \emph{Motor Imagery} (MI) is an intuitive, stimulus-independent approach: \textit{electroencephalography (EEG) sensors record brain activity while a person imagines moving a body part from a first-person perspective, without actual movement} (Figure~\ref{fig:scenarios})~\cite{marchesotti2016quantifying,lebon2010benefits,schuster2011best}. Unlike other BCI paradigms such as Event-Related Potentials (ERPs)~\cite{herrmann2001mechanisms} or Steady-State Visually Evoked Potentials (SSVEPs)~\cite{guo2022ssvep}, which depend on external cues and can induce fatigue, MI aligns naturally with volitional control and offers a more intuitive means of interaction. Beyond practical interfaces, MI studies also deepen our understanding of the neural mechanisms underlying motor and cognitive functions~\cite{munzert2009cognitive}.

A user's movement intention in MI arises from the synchronized firing of tens of thousands of neurons in motor cortical regions, generating neural source signals~\cite{tam2019human} with characteristic spatiotemporal patterns. As these signals travel toward the scalp, they traverse multiple biological layers, including the meninges, cerebrospinal fluid, skull, and scalp~\cite{bear2020neuroscience}. Each layer acts as a volume conductor and effective low-pass filter~\cite{kandel2000principles}, causing strong attenuation, spatial mixing, and additional distortion from noise (see \S\ref{pre:Principle}). As a result, the scalp-recorded EEG is a weak, distorted superposition of activities from multiple brain regions, which complicates accurate intent inference. Despite these complexities, recent approaches have harnessed machine learning to achieve encouraging results~\cite{zhang2018converting,altaheri2022physics,altaheri2023deep,ding2025eeg,an2023dual,gu2025cltnet,liao2025composite}. However, the wider adoption and practical deployment of the MI system remain difficult. Most existing approaches are designed as task-specific black-box models that lack a unified architectural foundation. This makes them fragile, struggling to aggregate knowledge from diverse datasets, adapt to variations in sensor configurations, or maintain performance in noisy, real-world conditions. Collectively, these limitations hinder the practical adoption of MI-based BCIs.

To fill this gap, we present \textbf{\systemname}, a new MI-based BCI framework that reliably and practically decodes imagined movements by addressing two core challenges:

\textbf{1) The prevailing paradigm of developing isolated, task-specific black-box models.} Most approaches employ independent, opaque models for each MI task. This fragmented, task-specific methodology lacks a unified architectural foundation. Consequently, these models are trained in isolation and fail to learn a robust and informative representation from the collective knowledge of diverse datasets. This not only leads to modest performance but also hinders interpretability, making it difficult to understand or trust the model's decisions.

\systemname addresses this by introducing a well-designed neural architecture for robust MI decoding. The main idea is to use the brain's forward signal pathway from cortex to scalp as a high-level design inspiration for our deep learning architecture, providing a structural rationale for our neural network components. Taking high-level design inspiration from the neural signal pathway~\cite{solodkin2004fine,bear2020neuroscience}, \systemname organizes decoding into three coupled modules. First, a \emph{signal filtering} stage mitigates noise and non-cortical interference. Second, a \emph{spatial representation learning} stage extracts robust spatial features from scalp recordings. Third, a \emph{feature aggregation and classification} stage processes these representations to infer the user's imagined movement, providing unified and interpretable decoding rather than task-specific black-box mappings.

\textbf{2) The inherent fragility of these models to electrode variability and signal noise.} Recent designs primarily target the EEG devices or data available to them, while real-world configurations vary in electrode number and placement, causing the model design limited to a given electrode configuration and the model trained on one configuration to fail being used when deployed on another. Furthermore, when MI-based BCI systems are deployed with mobile EEG devices, performance may degrade sharply due to the low SNR of EEG signals from commercial EEG devices.

To address the electrode heterogeneity, we introduce a \emph{spatially aware graph adapter} that projects inputs from arbitrary electrode configurations into a shared latent space. The adapter constructs a graph over electrodes based on their deployment coordinates (\eg 10–10 system positions~\cite{1010}) and learns layout-invariant embeddings that can gracefully handle missing channels, making the system to be applied to various EEG electrode configurations. To mitigate the low-SNR issue from consumer-grade mobile EEG, \systemname further employs \emph{multimodal auxiliary training}. During training, auxiliary streams, such as automatically generated visual cues, serve as privileged information that refines EEG representations and promotes noise-robust feature learning. At inference time, however, \systemname relies solely on EEG inputs. Together, these two designs directly address the challenges of electrode variability and signal noise, ensuring robust performance across heterogeneous devices and real-world environments.

We integrate above designs into \systemname, implement a prototype with a consumer EEG device Emotiv FLEX 2~\cite{Emotiv}, and deploy it on a Google Pixel 7 smartphone. This mobile deployment serves as a crucial validation of its real-world practicality, demonstrating that \systemname operates with exceptional efficiency on resource-constrained hardware. To comprehensively evaluate its performance, we conduct extensive experiments on three self-collected datasets (\systemname-DS-32C, 16C and 8C, gathered from 12 volunteers) and three public benchmark datasets (BCIC-2a~\cite{brunner2008bci}, BCIC-2b~\cite{leeb2008bci}, and MI-KU~\cite{lee2019eeg}). We compare \systemname with the recent methods including LGL-BCI~\cite{lu2025lgl}, CLTNet~\cite{gu2025cltnet}, and CIACNet~\cite{liao2025composite}. Overall, \systemname consistently maintains competitive performance across all six datasets. We also examine the performance of \systemname under various settings, including different user hair lengths, device wearing positions, and ambient noise levels, \etc Results show that \systemname maintains robust and reliable performance in diverse settings.

In summary, this paper makes the following key contributions:
\begin{itemize}
    \item We introduce \systemname, a deep learning architecture for MI decoding that takes high-level structural inspiration from the cortex-to-scalp signal pathway, employing three specialized neural modules (signal filtering, spatial representation learning, and feature classification) to yield robust decoding.
    \item We develop training mechanisms that make MI practical under real-world variability: a spatially aware graph adapter that handles heterogeneity in electrode settings (\ie numbers and placements) by projecting EEG data into a unified latent space, and multimodal auxiliary training that uses auxiliary cues only during training to learn noise-robust EEG features while requiring EEG alone at inference.
    \item We develop \systemname and conduct extensive evaluations on both consumer-grade and medical-grade datasets. Results show that \systemname consistently maintains competitive performance across all datasets.
\end{itemize}

%% file: Body/pre.tex
\section{Preliminary}
\label{pre}

\begin{figure}[t]
	\centering
	\includegraphics[width=2.8in]{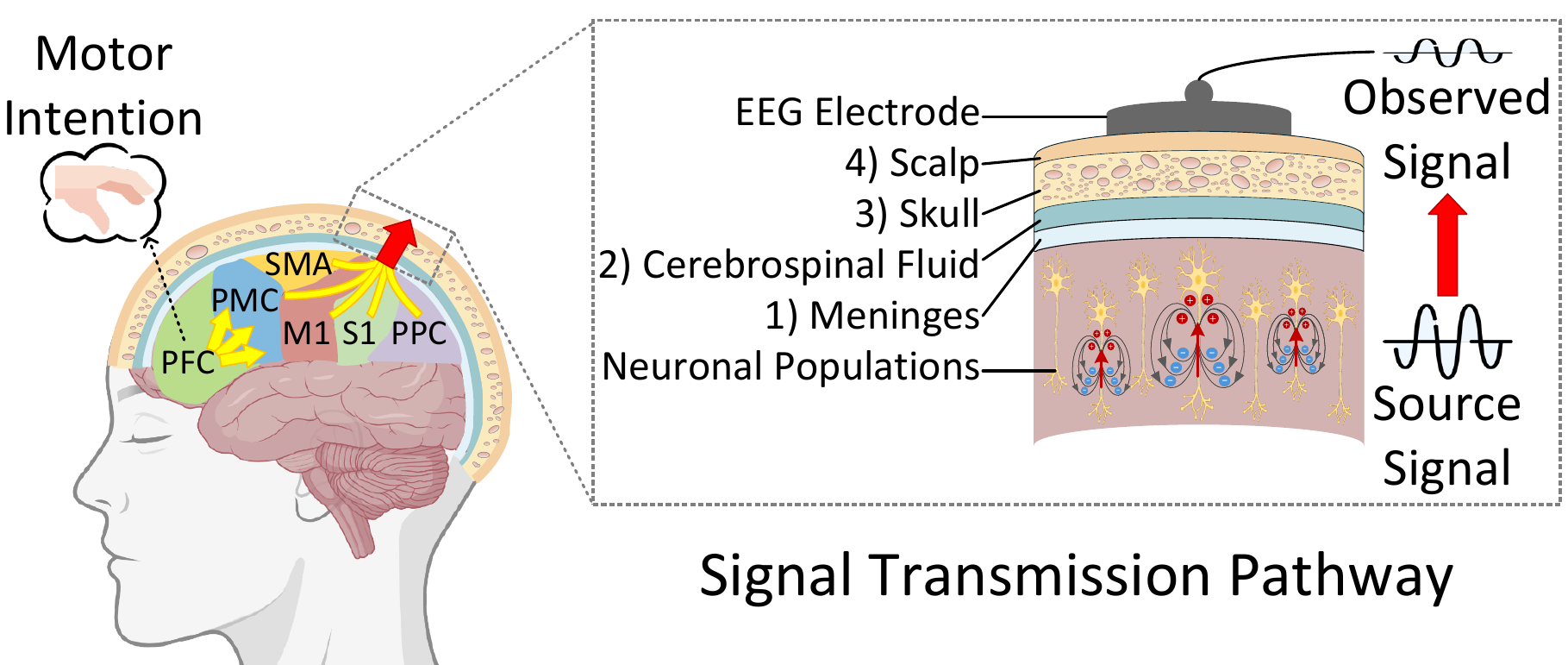}
    \caption{MI signal transmission pathway in the brain.}
	\label{fig:signal_path}
	\vspace{-.1in}
\end{figure}

\subsection{Basic Principles of Motor Imagery}
\label{pre:Principle}

MI refers to the mental rehearsal of a body movement from a first-person perspective without any actual physical execution.
To decode this mental activity accurately, it is important to understand how the neural signals are generated, how they are sensed, and what characteristics they exhibit, as these very principles give rise to the core challenges in practical MI-BCI systems.

\textbf{Generation and capture of neural signals.} This process begins when a person imagines a specific movement (\eg moving the left hand). Figure~\ref{fig:signal_path} illustrates the involved brain regions and the signal transmission pathway. The intention is first initiated in the Prefrontal Cortex (PFC), which then activates motor-related regions such as the Premotor Cortex (PMC), Supplementary Motor Area (SMA), and Primary Motor Cortex (M1). Neuronal populations in these regions fire \textit{synchronously}, generating neural \textbf{source signals}. As these source signals propagate toward the scalp, they pass through several physiological layers including the 1) meninges, 2) cerebrospinal fluid, 3) skull, and 4) scalp, where they undergo strong low-pass filtering and volume conduction effects~\cite{kandel2000principles} before reaching the surface. These faint activities can be captured non-invasively using EEG, which records electrical potentials from electrodes placed on the scalp.

This complex signal journey from a deep cortical source to the scalp is why MI decoding is so challenging. The resulting scalp EEG is a weak, distorted superposition of activities from multiple brain regions. This inherent complexity explains why treating the decoding process as a simple black-box problem often yields fragile and non-interpretable models; such models fail to account for the underlying physics of signal mixing and distortion.

\textbf{EEG rhythms \& event-related (de)synchronization.} EEG signals are commonly categorized into rhythms: $\delta$–$\theta$ (1–7 Hz) linked to deep sleep and drowsiness, $\alpha$ (8–12 Hz) associated with MI and relaxation, $\beta$ (13–30 Hz) related to active thinking and motor control, and $\gamma$ (>30 Hz) reflecting higher cognitive functions. Among these, while the $\alpha$ rhythm is most prominently associated with MI, other rhythms also contribute valuable information since MI engages multiple cognitive and sensorimotor processes. For example, when a person imagines clenching the left hand, the power of the $\alpha$ rhythm decreases, a phenomenon known as \textit{Event-Related Desynchronization (ERD)}. After the imagery task ends, the $\alpha$ power often rebounds, which is termed \textit{Event-Related Synchronization (ERS)}. Similar ERD and ERS phenomena are observed in the $\beta$ rhythm.

\begin{figure}[t]
	\centering
	\includegraphics[width=3in]{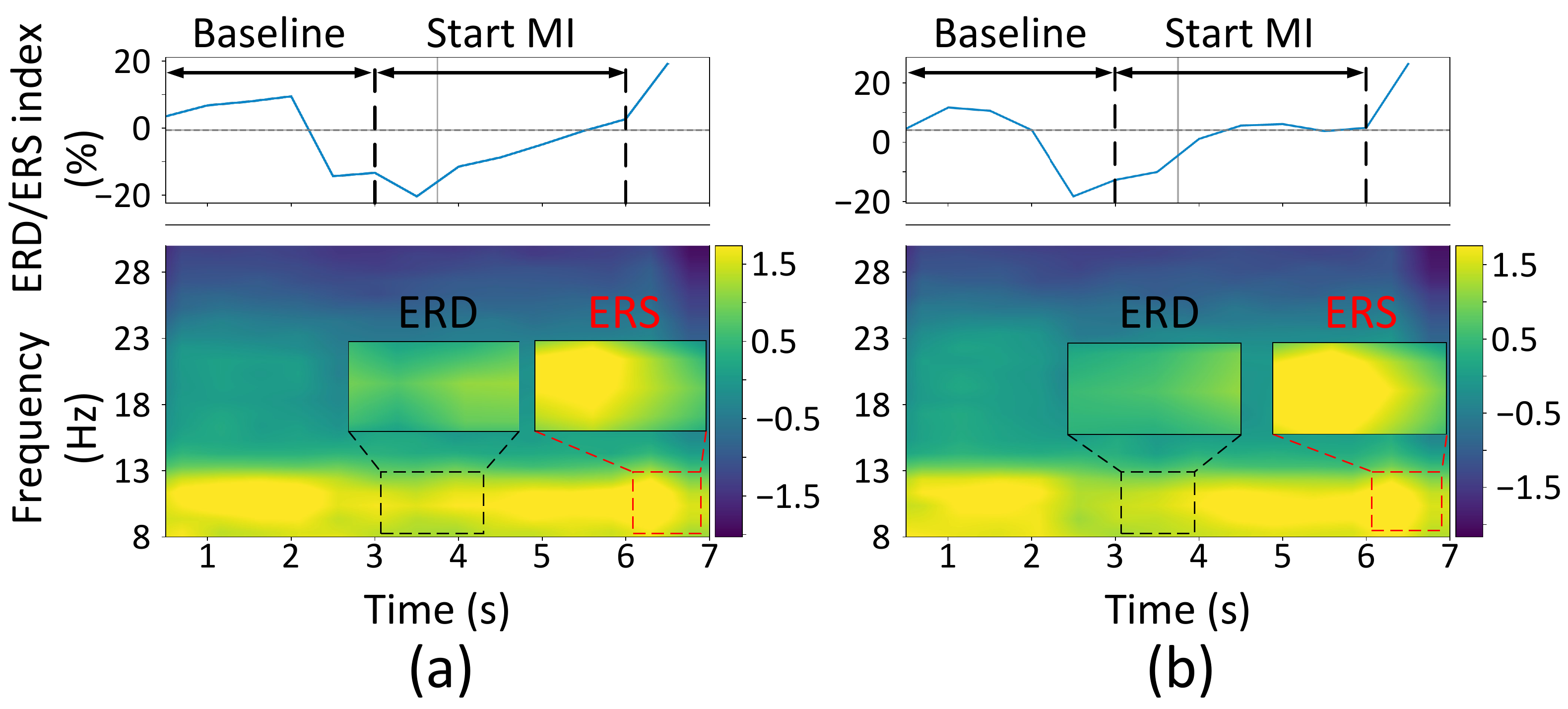}
	\vspace{-.1in}
        \caption{ERD and ERS patterns during motor imagery of clenching the (a) left hand and (b) right hand.}
	\label{fig:erd_ers}
	\vspace{-.22in}
\end{figure}

To quantify these changes, we use the \textbf{ERD or ERS index}, defined as the percentage change in band power relative to a baseline period~\cite{tang2016brain}, $\text{ERD/ERS index}~(\%) = \frac{A - R}{R} \times 100\%$, where $A$ is the power during the task and $R$ is the baseline power in the 8--30 Hz range (covering $\alpha$ and $\beta$ rhythms). Negative values indicate ERD (power decrease), while positive values indicate ERS (power increase) (Figure~\ref{fig:erd_ers}(a)). These repeatable EEG patterns provide the basis for inferring a user's intent, but they are often subtle and easily obscured by noise. In addition, imagery tasks for different limbs (\eg imagining clenching the right hand versus the left hand) produce highly similar signal patterns (Figure~\ref{fig:erd_ers}(b), imagining clenching the right hand), making them nontrivial to distinguish.

\subsection{Application Scenarios}
\label{pre:app}

Leveraging its non-invasive and intuitive nature, MI holds great potential in healthcare and daily interaction.

\parahead{Limb replacement and rehabilitation.} For patients with severe motor impairments, MI acts as a critical bridge between intention and action~\cite{saruco2024towards,liao2023motor}. It enables assistive control (\eg wheelchair operation~\cite{palumbo2021motor}) and allows clinicians to objectively monitor neural recovery even before physical function is regained~\cite{saruco2024towards}.

\parahead{Human–computer interaction.} MI offers a silent, hands-free, and privacy-preserving control channel for the broader population~\cite{nwagu2023eeg}. It facilitates seamless interaction in smart homes~\cite{zhuang2020brain} and provides an additional intuitive input modality for hands-busy scenarios like virtual reality and gaming~\cite{arpaia2024endless}.

\subsection{System Overview}
\label{pre:overview}

\begin{figure}[b]
	\centering
	\includegraphics[width=2.8in]{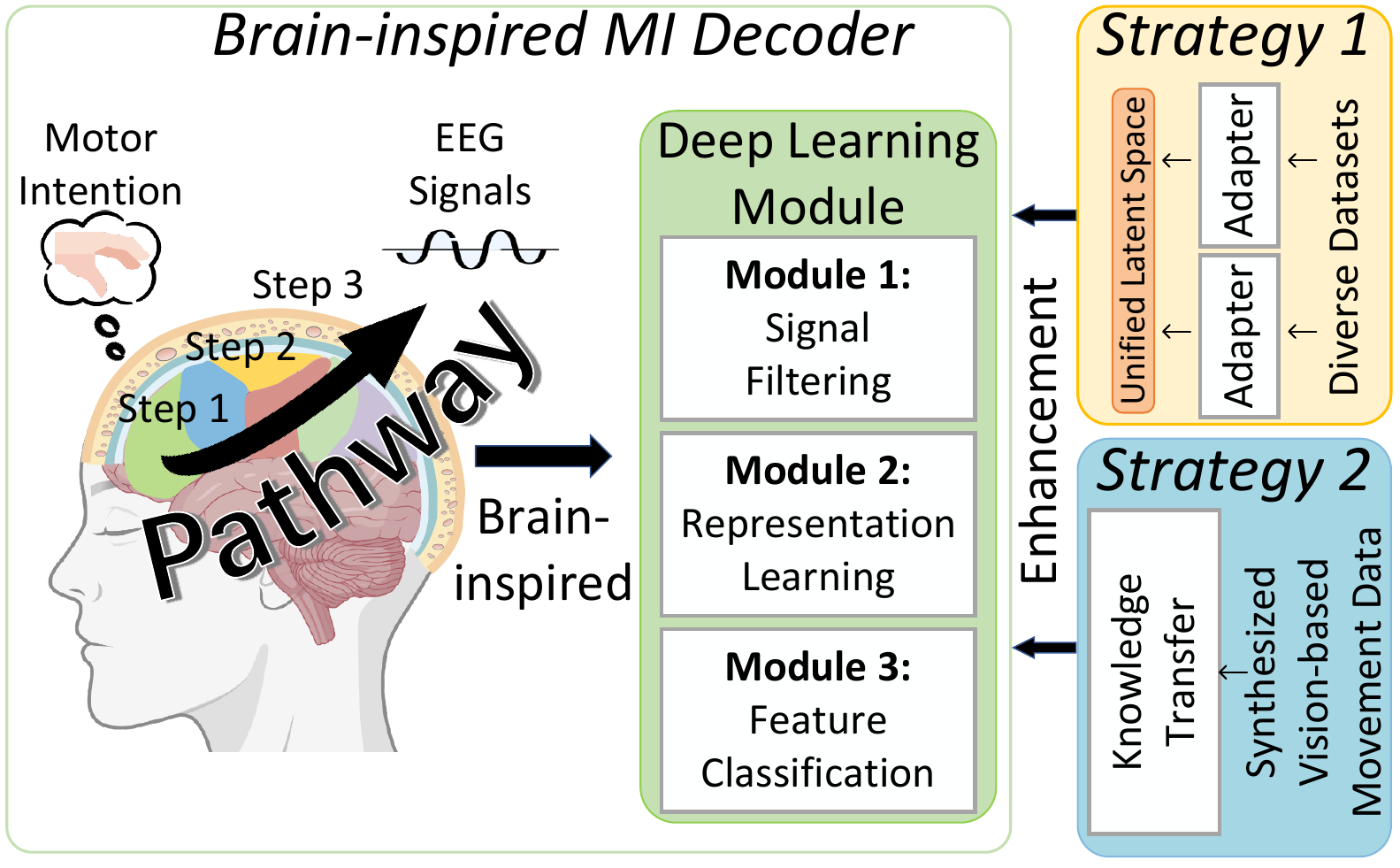}
	\vspace{-.1in}
    \caption{Overview of the \systemname design.}
	\label{fig:overview}
	\vspace{-.2in}
\end{figure}

Figure~\ref{fig:overview} shows the overview of \systemname, which consists of two main components.

\parahead{1) Brain-inspired MI decoder (\S\ref{design:decoder}).} 
The core of \systemname is a decoder architecture inspired by the brain's forward signal generation process. It draws structural inspiration from this pathway through three deep learning modules: signal filtering, spatial representation learning, and feature aggregation and classification. This design improves decoding accuracy, interpretability, and reliability, providing a unified architectural foundation for MI decoding.

\parahead{2) Practical training framework (\S\ref{design:enhancement}).}  
To address the configuration heterogeneity and low-quality EEG signal issues, we further propose two complementary training strategies that enhance the MI decoder proposed in \S\ref{design:decoder}. First, a spatially-aware graph adapter addresses heterogeneity in electrode settings (\ie varying numbers and placements) by projecting EEG data into a unified latent space. This makes the decoder agnostic to the input electrode configuration, achieved by leveraging multiple diverse datasets as robust priors during training. Second, a cross-modality knowledge transfer framework leverages synthesized vision-based movement data to guide training, enabling the decoder to learn more discriminative and robust motor features.

%% file: Body/design.tex
\section{Brain-inspired MI Decoder}
\label{design:decoder}

\subsection{Design Rationale}
The core idea of our decoder is to address the inherent complexity of how motor intentions are transformed into scalp-recorded EEG signals. A user's motor intention, denoted as $x_{thought}$, is not directly observable on the scalp. Instead, it is first encoded by neuronal populations into neural source signals $\hat{S}$. These source signals then propagate through multiple biological layers, where they are attenuated and spatially mixed, resulting in the propagated scalp signal, denoted as $\hat{X}_{prop}$. Finally, this signal is contaminated by various external and physiological noise sources, producing the observed low-SNR EEG signal $X$:
\begin{equation*}
    x_{thought} \rightarrow \hat{S} \rightarrow \hat{X}_{prop} \rightarrow X.
\end{equation*}

Decoding thus becomes a challenging \textit{ill-posed inverse problem}: how to accurately infer $x_{thought}$ from the noisy EEG $X$. While one could attempt to model and invert the entire biophysical process exactly, this is infeasible due to the incomplete understanding of brain mechanisms and the prohibitive computational cost.

Instead, we draw inspiration from the \textit{encoder–decoder} paradigm in deep learning. The brain's forward generation of signals (from $x_{thought}$ to $X$) can be regarded as a complex ``encoder'' governed by neurophysiological dynamics. Our idea is to construct a deep learning architecture inspired by these key stages, utilizing standard neural network components to process the data, rather than attempting a literal biological inversion. By using these physiological stages as high-level architectural priors, \systemname is guided toward more effective representation learning and robust decoding.

\begin{figure}[t]
	\centering
	\includegraphics[width=3in]{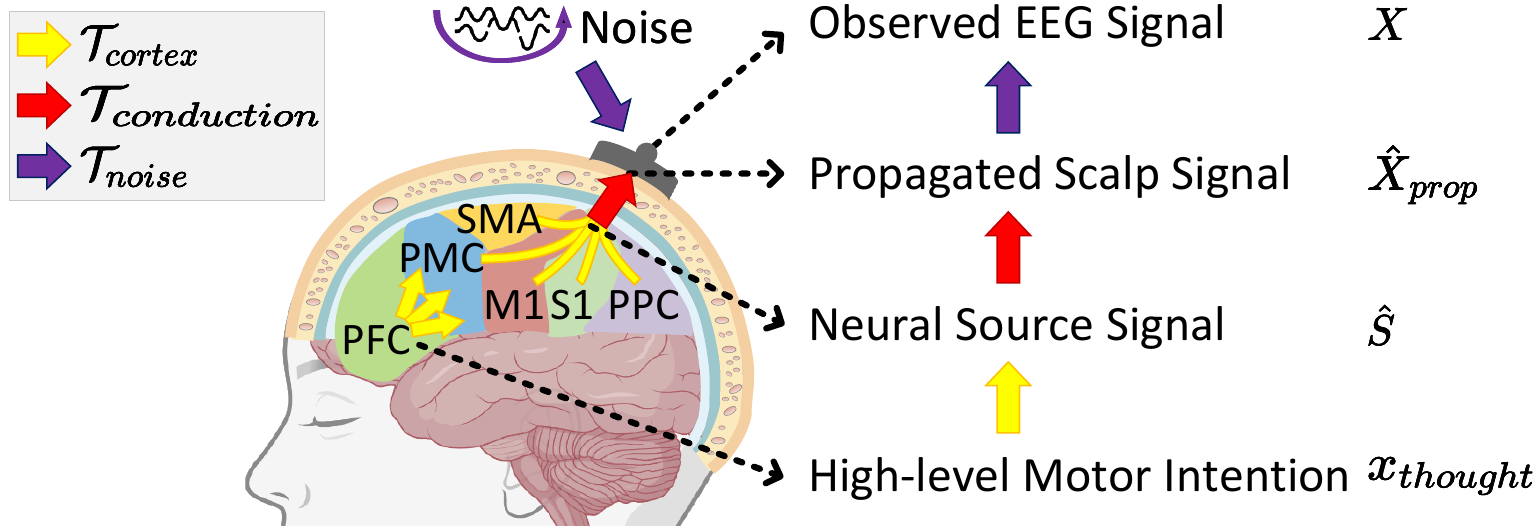}
	\vspace{-.05in}
        \caption{Illustration of the forward generation process.}
	\label{fig:decoder}
	\vspace{-.15in}
\end{figure}

\subsection{The Forward Generation Process}
To guide the deep learning architecture design, we conceptually abstract the forward generation of EEG signals from $x_{thought}$ to $X$ as a high-level composite function consisting of three stages (Figure~\ref{fig:decoder}):
\begin{equation}
    X = [\mathcal{T}_{noise} \circ \mathcal{T}_{conduction} \circ \mathcal{T}_{cortex}](x_{thought}),
\end{equation}
where $\circ$ denotes function composition.

\parahead{1) Cortical processing transform $\mathcal{T}_{cortex}$:} 
\begin{equation}
    \hat{S} = \mathcal{T}_{cortex}(x_{thought}).
    \label{eq:transform1}
\end{equation}

This transform models the activity of motor-related brain regions such as the PMC, SMA, and M1 shown in Figure~\ref{fig:decoder}. It converts the high-level motor intention $x_{thought}$, initiated in the PFC, into multichannel neural source signals $\hat{S}$ with characteristic spatiotemporal patterns~\cite{solodkin2004fine}. These signals arise from the synchronized firing of neural ensembles in differential rhythmic patterns.

\parahead{2) Volume conduction transform $\mathcal{T}_{conduction}$:}
\begin{equation}
    \hat{X}_{prop} = \mathcal{T}_{conduction}(\hat{S}).
    \label{eq:transform2}
\end{equation}

As $\hat{S}$ propagates outward, it passes through the meninges, cerebrospinal fluid, skull, and scalp. These biological layers act as low-pass filters and volume conductors~\cite{kandel2000principles}, attenuating the signals and mixing them across electrodes. The result is $\hat{X}_{prop}$, a blurred linear superposition of multiple neural sources recorded at the scalp.

\parahead{3) Noise contamination transform $\mathcal{T}_{noise}$:} 
\begin{equation}
    X = \mathcal{T}_{noise}(\hat{X}_{prop}).
    \label{eq:transform3}
\end{equation}

Finally, the scalp-level signals are contaminated by multiple sources of noise~\cite{li2021review}, including electromagnetic interference, device-related electronic noise, and physiological artifacts such as eye blinks or muscle activity. The observed EEG $X \in \mathbb{R}^{C \times T}$ is therefore a low-SNR signal, where $C$ is the number of channels and $T$ is the number of time samples.

\begin{figure}[t]
	\centering
	\includegraphics[width=3in]{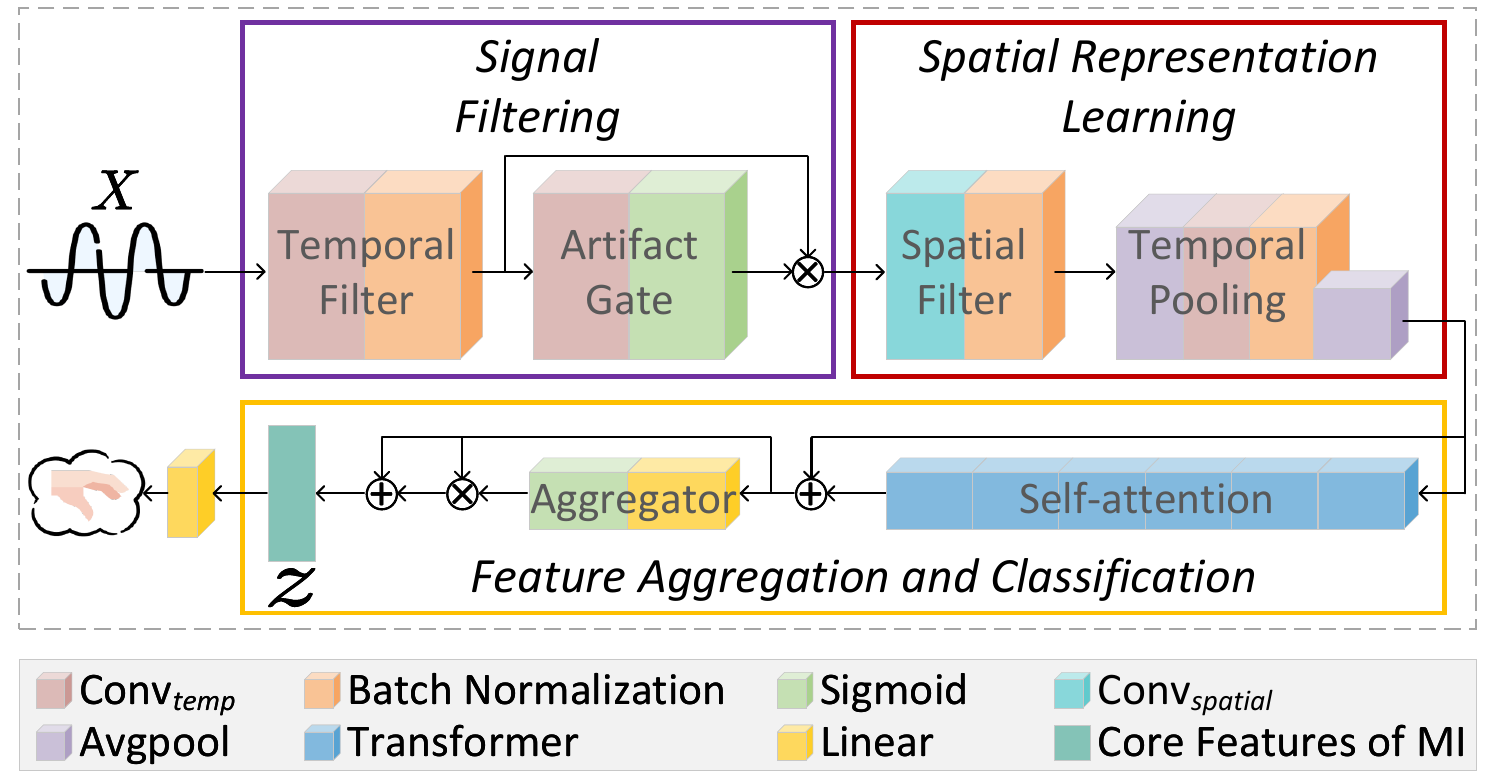}
	\vspace{-.1in}
        \caption{Brain-inspired MI decoder architecture.}
	\label{fig:decoder_arch}
	\vspace{-.1in}
\end{figure}

\subsection{Brain-inspired MI Decoder Architecture}

Taking inspiration from the forward generation model, our MI decoder is designed with neural network modules that conceptually align with these physiological stages, mapping the noisy EEG sequence to a classification output. As shown in Figure~\ref{fig:decoder_arch}, the decoding pipeline transforms the EEG input $X$ into a compact feature vector $z$ for classification. It consists of three modules, each drawing structural inspiration from a specific stage in the forward process.

\parahead{1) Signal filtering module.}
Conceptually addressing the noise contamination stage, this module employs learnable temporal convolutions and gating mechanisms to mitigate noise and artifact contamination from the raw EEG $X$. It includes two sub-modules:

\textbf{Learnable temporal filter.}  
EEG signals contain both informative rhythms and a large amount of task-irrelevant activity across frequencies, making filtering essential to highlight MI–related bands (\eg $\alpha$ and $\beta$ rhythms) while suppressing noise and unrelated components. Convolution is mathematically equivalent to finite impulse response (FIR) filtering, with the convolution kernel serving as the filter coefficients. Therefore, a temporal convolution can naturally function as a bank of parallel, learnable FIR filters, where larger kernels correspond to longer filter windows in the time domain. Instead of relying on manually defined passbands, the network learns optimal filter weights $W_{temp}$ directly from data:
\begin{equation}
X_{filtered} = Conv_{temp}(X) = W_{temp} * X.
\end{equation}

\textbf{Adaptive artifact suppression gate.}  
Even after temporal filtering, EEG often contains burst-like, non-stationary artifacts such as eye blinks, jaw clenching, or head movements. These artifacts typically appear with large amplitudes that can easily overshadow the subtle MI-related rhythms if not explicitly addressed. To mitigate this, we introduce an adaptive gating mechanism that learns to detect and suppress such artifacts. Specifically, a convolution followed by sigmoid activation generates a dynamic mask $G \in [0, 1]^{C \times T}$ that is synchronized with the signal:
\begin{equation}
G = \sigma(Conv_{gate}(X_{filtered})).
\end{equation}

Values near 0 indicate corrupted segments to be down-weighted, while values near 1 preserve clean portions of the signal. This soft, element-wise weighting avoids hard signal removal and instead provides a fine-grained, data-driven suppression of artifacts.  

The propagated scalp signal is then obtained through:
\begin{equation}
\hat{X}_{prop} = G \odot X_{filtered},
\end{equation}
where $\odot$ denotes element-wise multiplication.

\parahead{2) Spatial representation learning module.}
Conceptually addressing the volume conduction effect, this module uses spatial convolutional filters to learn robust, localized representations from the filtered scalp signals and extract key rhythmic features.

\textbf{Spatial filter.}
Because each electrode records a superposition of activities from multiple cortical regions, direct use of scalp potentials makes it difficult to isolate MI-related signals. To address this, we apply a spatial convolution across all electrode channels that learns weights $W_{spatial}$ to optimally recombine the mixed signals. This operation acts as a data-driven spatial filter, similar in spirit to methods such as CSP (Common Spatial Patterns)~\cite{lotte2010regularizing}, but optimized end-to-end within the network. By emphasizing discriminative spatial patterns, it separates overlapping sources and learns localized spatial representations $\hat{S}$:
\begin{equation}
\hat{S} = Conv_{spatial}(\hat{X}_{prop}) = W_{spatial} \cdot \hat{X}_{prop}.
\end{equation}

\textbf{Temporal pooling.}
Even after spatial convolution, the intermediate representations $\hat{S}$ remain continuous time-series signals that need to be characterized in terms of their oscillatory content. MI is known to modulate specific neural rhythms (\eg $\alpha$ and $\beta$), so extracting these rhythmic features is crucial for distinguishing different MI tasks. To this end, we apply temporal convolutions followed by average pooling, which together act as learnable band-pass filters and summarizers. The convolution learns frequency-selective patterns, while pooling aggregates them into compact feature tokens. The result is a set of discriminative rhythm representations $F_{rhythm}$:
\begin{equation}
F_{rhythm} = AvgPool(Conv_{rhythm}(\hat{S})).
\end{equation}

These features capture the essential oscillatory signatures of MI while reducing redundancy and noise, preparing them for high-level inference in the next stage.

\parahead{3) Feature aggregation and classification module.}
Rather than processing raw signals, this final module performs classification based on the rhythmic features $F_{rhythm}$ extracted by the previous layers.

\textbf{Self-attention module.}
MI does not arise from a single brain region but from the coordinated activity of several sensorimotor areas working together. To correctly decode intention, it is therefore necessary to capture not only the strength of individual rhythms but also the interactions between them.  

To achieve this, we use Transformer layers. A Transformer’s self-attention mechanism allows each feature token to dynamically weigh and relate to all others, regardless of their spatial or temporal distance. This ability to model long-range dependencies makes it well suited to represent the synergy across distributed brain regions. The output of the Transformer is then combined with the original rhythm features $F_{rhythm}$ through a residual connection. This creates an enriched set of features $F_{enhanced}$ that benefits from both the local rhythmic patterns and their global interactions:
\begin{equation}
F_{enhanced} = F_{rhythm} + Transformer(F_{rhythm}).
\end{equation}

\textbf{Intention feature aggregator.}
Not all features contribute equally to distinguishing motor intentions. To emphasize the most informative components, we introduce a gating mechanism implemented with a linear layer followed by a sigmoid activation. The gate $G_{agg}$ selectively weights the enhanced features through element-wise multiplication. The resulting sequence is then flattened into a compact vector $z$ that captures the user’s motor intention.
\begin{align}
G_{agg} &= \sigma(Linear_{gate}(F_{enhanced})), \\
z &= Flatten(G_{agg} \odot F_{enhanced} + F_{enhanced}).
\end{align}

Finally, the vector $z$ is projected by a linear classifier to predict the motor intention $x_{thought}$. The decoder is trained end-to-end with a cross-entropy loss, encouraging the network to learn a feature space in which different motor intentions are clearly separable.

\begin{table}[t]
\centering
\caption{Comparison of key metrics between medical-grade and consumer-grade devices.}
\label{tab:device_comparison}
\begin{tabular}{ccc}
\toprule
\textbf{Parameter} & \textbf{Medical} & \textbf{Consumer} \\
\midrule
\textbf{Electrode Type} & Ag/AgCl Wet &  Dry or Saline\\
\textbf{Data Resolution} & 24-bit & 16-bit \\
\textbf{Voltage Sensitivity} & 0.39 nV & 510 nV \\
\textbf{Sampling Rate} & High & Low \\
\textbf{Channel Density} & Dense & Sparse \\
\textbf{Price (USD)} & \$30,000+ & \$500 - \$2,000 \\
\bottomrule
\end{tabular}
\end{table}

\section{Practical Training Framework for Robust MI EEG Decoding}
\label{design:enhancement}
\subsection{Design Motivation}

The MI decoder proposed in \S\ref{design:decoder} improves both accuracy and interpretability, providing a unified architectural foundation. However, for real-world deployment, we need to further address two practical challenges that hinder the widespread adoption of MI-BCIs.

First, EEG data is plagued by heterogeneity in electrode numbers and placements. This poses two critical problems. On one hand, it obstructs knowledge aggregation from diverse datasets, which are invaluable for training a powerful model but feature disparate electrode layouts. On the other hand, it creates a critical gap: a model trained for a specific electrode layout typically fails to work when deployed on a system with a different configuration. This fundamental mismatch between data sources, both across training sets and between training and deployment, severely limits a model's scalability and real-world applicability.

Second, non-intrusive EEG recordings suffer from a low-SNR, a problem that is especially severe in consumer-grade devices. As shown in Table~\ref{tab:device_comparison}, compared with their high-cost, medical-grade counterparts, these consumer-grade devices typically use dry or saline electrodes, have lower resolution and sensitivity, and provide fewer recording channels. These limitations greatly weaken the captured MI-related rhythms: subtle neural oscillations may be attenuated below detection thresholds, while sparse spatial sampling blurs activity from distinct cortical regions. As a result, the signals collected by consumer devices have much lower SNR, making them significantly harder to decode reliably.

\subsection{Addressing Electrode Heterogeneity}
\label{design:enhancement:same}

To address the challenge of electrode heterogeneity, we introduce a novel component designed to make our model agnostic to specific electrode layouts: the spatially-aware graph adapter. Its primary function is to transform EEG data from arbitrary layout into a unified, standardized representation. This approach directly solves the deployment mismatch problem and paves the way for aggregating knowledge from diverse data to further enhance model robustness.

\parahead{1) Spatially-aware graph adapter.}
To make our model functional across diverse electrode configurations, we introduce a spatially-aware graph adapter in \systemname.

\parahead{1.1) Insights.}
The spatially-aware graph adapter transforms heterogeneous EEG inputs of size $(C, T)$ into a unified tensor $(C_{fix}, T_{fix})$, enabling consistent processing by the shared MI decoder proposed in §\ref{design:decoder}. In doing so, it not only makes pre-training on heterogeneous datasets feasible but also allows the learned features to transfer seamlessly to varied devices. Its design is guided by two insights:

\begin{figure}[t]
    \centering
    \includegraphics[width=3in]{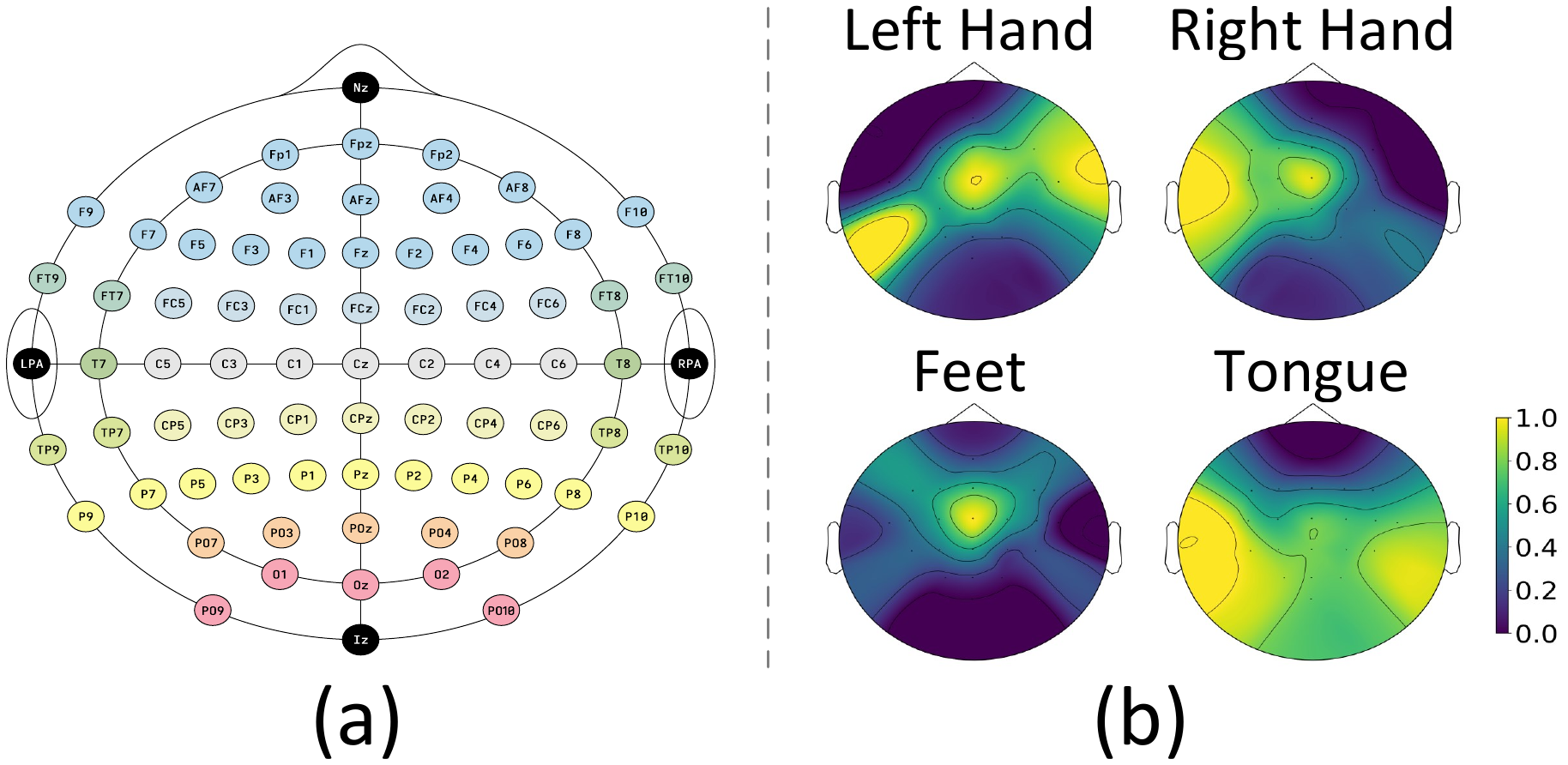}
    \vspace{-.1in}
    \caption{(a) International 10-10 electrode system. (b) Task-dependent importance of different channels, based on average signal amplitude in MI tasks.}
    \label{fig:obser}
    \vspace{-.15in}
\end{figure}

\textit{Structured spatial prior.}  
The international 10–10 system~\cite{1010} provides a standardized map of electrode positions across the scalp, as shown in Figure~\ref{fig:obser}(a). Different devices or datasets typically record from different subsets of these electrodes, leading to heterogeneity in channel numbers and placements. To handle this, we model electrodes as nodes in a graph, with edges encoding physical distances between locations. This representation preserves the true spatial layout of electrodes and enables consistent integration across datasets, rather than treating channels as an unordered list.  
    
\textit{Dynamic channel importance.}  
Even within the same electrode layout, not all electrodes contribute equally to every motor imagery task. For instance, hand motor imagery produces contralateral brain activity: imagining movement of the left hand primarily activates electrodes on the right side of the scalp, whereas imagining movement of the right hand primarily activates those on the left, as illustrated in Figure~\ref{fig:obser}(b). Similar task-specific patterns can be observed for other imagery tasks such as feet or tongue movements. To capture this variability, we employ a graph attention mechanism that adaptively assigns weights to electrodes based on task relevance. This allows the model to focus on the most informative channels while down-weighting irrelevant ones.  

\parahead{1.2) Design.}  
Building on the above insights, the adapter works in three steps for each heterogeneous dataset:   

\textbf{Step 1: Graph initialization.}  
Each electrode is treated as a graph node. We assign initial node features based on signal statistics, including band power in $\delta$–$\theta$ (1–7 Hz), $\alpha$ (8–12 Hz), $\beta$ (13–30 Hz), and $\gamma$ (31–45 Hz) rhythms, together with signal variance. At the same time, we construct the graph structure using the standardized 10–10 electrode coordinates: pairwise Euclidean distances between electrodes are converted into an adjacency matrix via a Gaussian kernel, which encodes spatial proximity as graph edges.  

\textbf{Step 2: Attention-based aggregation.}  
The initialized graph and raw EEG signals are processed by a Graph Attention Network (GAT). The GAT learns to dynamically weight information from different electrodes, amplifying task-relevant channels while suppressing less informative ones. This produces an enriched feature representation reflecting both spatial structure and task relevance.

\textbf{Step 3: Unified projection.}  
Since different datasets have varying channel numbers and locations, the aggregated features are finally mapped to a fixed channel dimension $C_{fix}$ (32 in our implementation) through a linear layer. The temporal dimension $T$ is simultaneously aligned to $T_{fix}$ using interpolation. This ensures all datasets are represented in a consistent tensor shape $(C_{fix}, T_{fix})$ for subsequent decoding.  

The above steps transform heterogeneous EEG data $(C, T)$ into a unified and informative representation $(C_{fix}, T_{fix})$ that preserves spatial structure and task-relevant information, ready for processing by the brain-inspired MI decoder. 

\begin{figure}[t]
    \centering
    \includegraphics[width=3.2in]{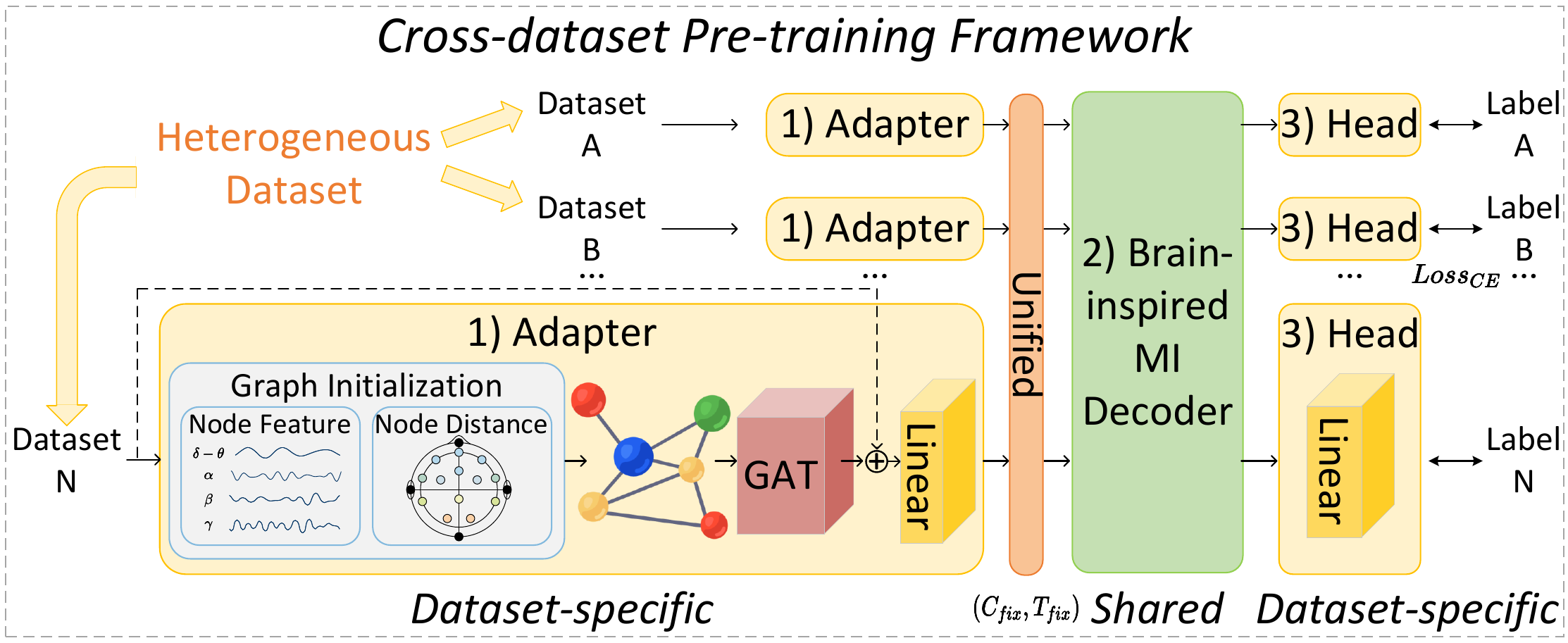}
    \vspace{-.1in}
    \caption{Cross-dataset pre-training framework integrating heterogeneous datasets through dataset-specific adapters and heads, with a shared MI decoder.}
    \label{fig:pretrain}
    \vspace{-.15in}
\end{figure}

\parahead{2) Cross-dataset pre-training framework.}  
With the proposed adapter, MI model can now process data from any electrode configuration in 10-10 system, resolving the critical challenge of electrode heterogeneity. This capability unlocks a powerful secondary benefit: we can now train a single, robust model by aggregating knowledge from multiple heterogeneous datasets simultaneously. To leverage this, we design a cross-dataset pre-training framework, illustrated in Figure~\ref{fig:pretrain}, which enhances the model's feature representations by exposing it to a wide variety of data. The pipeline consists of three main components: 

\textit{Dataset-specific adapters.} Each heterogeneous dataset is first processed by its own adapter, which converts raw EEG into the unified representation $(C_{fix}, T_{fix})$ while preserving spatial structure and task-relevant information.  

\textit{Shared MI decoder.} The unified features are then passed into the brain-inspired MI decoder introduced in §\ref{design:decoder}. This serves as a common backbone, enabling the model to learn shared neural feature priors across datasets.  

\textit{Dataset-specific heads.} Finally, the shared features are mapped to the label space of each dataset by a lightweight classification head, ensuring compatibility with different experimental tasks and label definitions.   

Overall, this design offers a comprehensive solution to the challenge of electrode heterogeneity. First and foremost, the spatially-aware graph adapter solves the critical problem of deployment mismatch, making the model inherently flexible and scalable to new datasets and hardware configurations. Building on this foundation, the cross-dataset pre-training framework further enhances model robustness and performance. By learning from diverse data sources, the shared decoder acquires rich, transferable neural feature priors. We currently employ a sequential dataset iteration strategy. Specifically, the model iterates through the full content of each dataset sequentially within a single training epoch.

\subsection{Addressing Low Signal Quality}
\label{design:enhancement:different}
 
To address the challenge of low signal quality, we introduce an auxiliary modality with clearer and more discriminative features to guide the fine-tuning phase after pre-training with §\ref{design:enhancement:same}. Rather than relying on additional sensors, this modality is generated through automated skeleton data synthesis. The key idea is that skeleton data provides a high-level structural representation of the instructed movement: it is inherently noise-free, requires minimal effort to generate with modern vision techniques, and captures distinct dynamic patterns for different imagined actions. As such, the skeleton modality serves as a low-cost but effective ``teacher'', guiding the MI decoder (``student'') through cross-modality knowledge transfer.

\begin{figure}[t]
	\centering
	\includegraphics[width=3.2in]{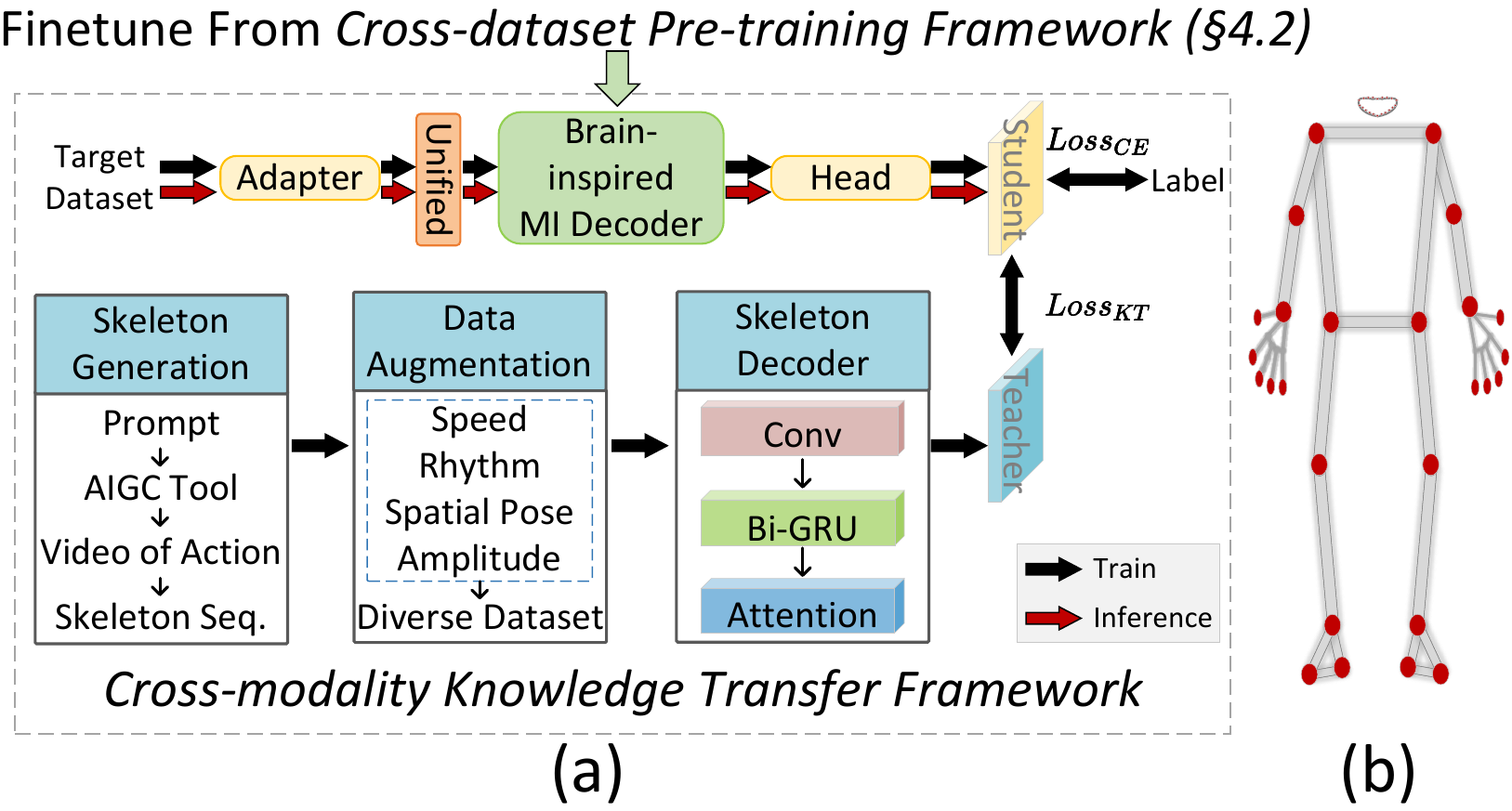}
	\vspace{-.1in}
    \caption{(a) Cross-modality knowledge transfer framework. (b) Skeleton keypoints extracted by MediaPipe.}
	\label{fig:skeleton}
	\vspace{-.1in}
\end{figure}

\parahead{1) Automated skeleton data generation.}  
To obtain the skeleton modality without extra sensors, we leverage AIGC tools (\eg vivago.ai~\cite{ai}) to generate short videos of target MI actions. Given a frontal photo (optional) and a text prompt describing the action, the tool synthesizes the video, from which we extract 3D coordinates of 46 body, hand, and face keypoints using Google MediaPipe~\cite{MediaPipe} (Figure~\ref{fig:skeleton}(b)). This forms a high-quality skeleton sequence for each action. This incurs only a one-time, low-effort generation overhead. Because these skeleton sequences are generated from static text descriptions rather than recorded simultaneously with the user's EEG, they share the same total duration but do not require strict frame-by-frame temporal alignment.

\parahead{2) Data augmentation pipeline.}  
Since one generated skeleton sequence is insufficient for robust training, we expand it into a diverse dataset by simulating natural variations in how people repeatedly perform the same action. This is achieved through four types of augmentations:  

\begin{itemize}
    \item \textbf{Speed variation:} the sequence is resampled at different rates to mimic actions performed faster or slower, reflecting natural differences in execution speed.  
    \item \textbf{Rhythm variation:} we apply non-linear temporal warping so that different phases of the action (\eg onset, peak, and offset) take variable durations, simulating irregular or personalized rhythms.  
    \item \textbf{Spatial variation:} small random global rotations are applied to the 3D coordinates to account for posture changes such as leaning forward, which do not alter the core action.  
    \item \textbf{Amplitude variation:} all coordinates are scaled by random factors to simulate changes in movement size, such as clenching more tightly or moving with smaller gestures.  
\end{itemize}

\parahead{3) Cross-modality knowledge transfer framework.}  
We design a skeleton decoder as the teacher network (Figure~\ref{fig:skeleton}(a)). It combines temporal convolutions for local motion patterns, a bidirectional GRU for long-range dependencies, and multi-head attention to emphasize key frames, producing compact and discriminative features. The MI decoder (student) is trained jointly with the skeleton decoder, using a composite objective:  
\begin{equation*}
    Loss_{total} = Loss_{CE} + Loss_{KT},
\end{equation*}  
where $Loss_{CE}$ is the cross-entropy loss with ground-truth labels, and $Loss_{KT}$ is the KL divergence that encourages the MI decoder to align its predictions with those of the skeleton decoder. This knowledge transfer allows the MI decoder to benefit from the clearer skeleton modality during training.   
 
\parahead{4) Inference.}  
After training, only the MI decoder is retained for inference. It benefits from the transferred skeleton knowledge and achieves stronger decoding performance on low quality EEG signals, without requiring skeleton data at runtime.

%% file: Body/evaluation.tex
\section{Evaluation}
\label{exp:eval}

\subsection{Experimental Setup}
\label{exp:setup}

\begin{figure}[h]
	\centering
	\includegraphics[width=3in]{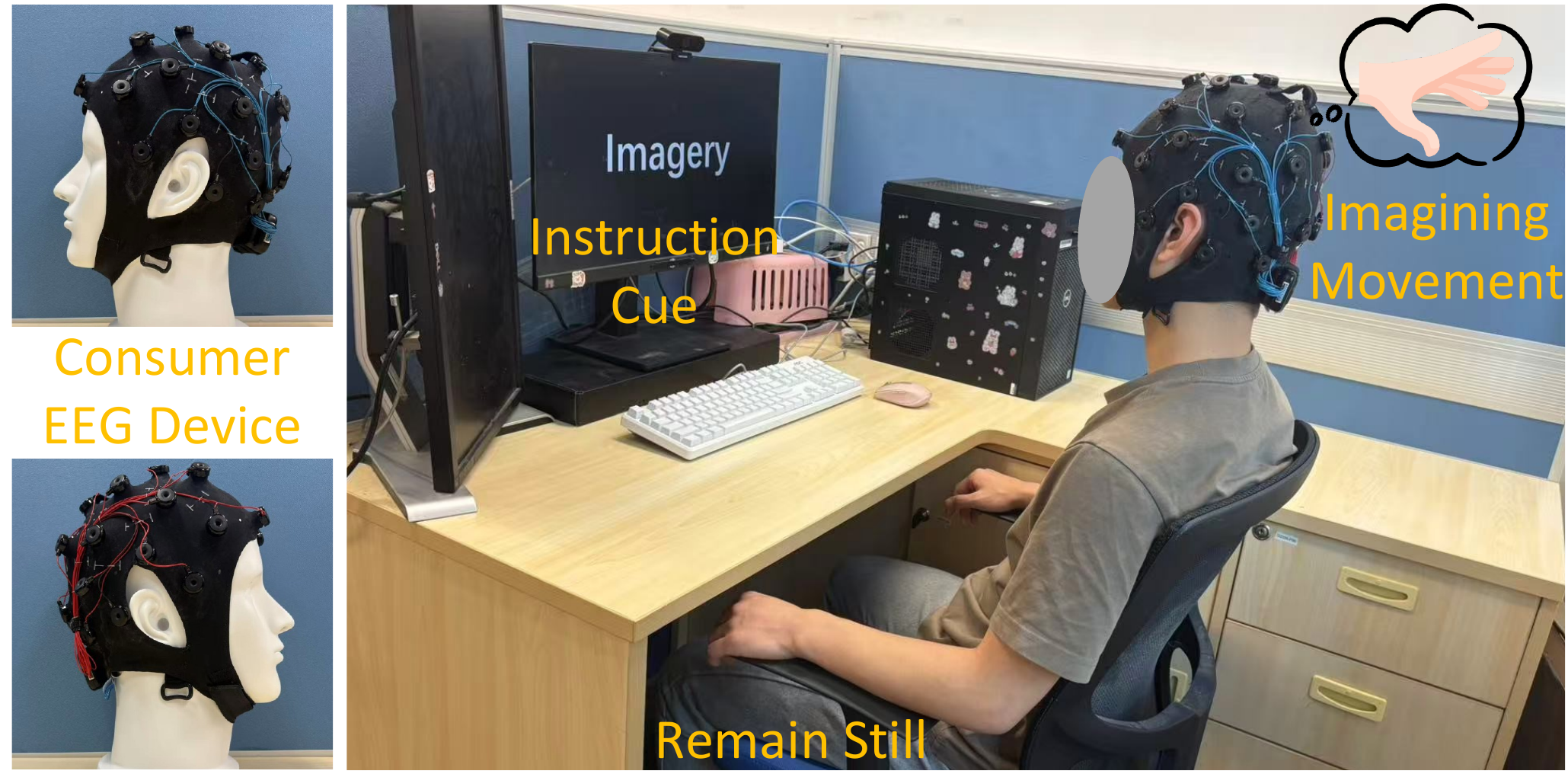}
	\vspace{-.15in}
        \caption{Illustration of the experimental setup.}
	\label{fig:exp_scenario}
	\vspace{-.1in}
\end{figure}

\parahead{Implementation.} \systemname is built on a consumer-grade headset Emotiv FLEX 2~\cite{Emotiv} as shown in Figure~\ref{fig:exp_scenario}, and the model is implemented using Python 3.8 and PyTorch 1.12. The model training follows a two-stage procedure. We use the Adam optimizer for all training stages. First, we employ the cross-dataset pre-training framework (\S\ref{design:enhancement:same}) to pre-train the shared MI decoder (\S\ref{design:decoder}) on heterogeneous public, medical-grade datasets (detailed below). This pre-training phase runs for 1000 epochs with a batch size of 16 and a learning rate of 0.001. After pre-training, the model weights, rich with prior knowledge, are saved. Subsequently, we use the cross-modality knowledge transfer framework (\S\ref{design:enhancement:different}) to fine-tune the pre-trained model on the target consumer-grade, low-quality dataset. The training parameters for this stage are kept consistent with the pre-training phase, with 1000 epochs, a batch size of 16, and a learning rate of 0.001, respectively. All models are trained on an NVIDIA V100 GPU. After training, we deploy the model to a Google Pixel 7 phone using Pytorch Mobile for inference to validate its practical performance in a mobile environment.

\begin{table*}[t]
\centering
    \caption{Summary of datasets used for evaluation. The first three are self-collected under different channel configurations, while the last three are public benchmark datasets.}
    \label{tab:datasets}
    
    \begin{tabularx}{\textwidth}{@{} c c c C c C c C @{}}
        \toprule
        \textbf{\multirow{2}{*}{Dataset}} & 
        \textbf{\multirow{2}{*}{Subjects}} &
        \textbf{\multirow{2}{*}{Channels}} &
        \textbf{\multirow{2}{*}{Device}} & 
        \textbf{\multirow{2}{*}{Electrode Type}} & 
        \textbf{Resolution (Sensitivity)} & 
        \textbf{\multirow{2}{*}{Cost (USD)}} & 
        \textbf{Collection Time Per User} \\
        \midrule
        
        NeuroPath-DS-32C & 12 & 32 & \multirow{3}{*}{\begin{tabular}{@{}c@{}}Consumer-grade \\ (Emotiv FLEX 2)\end{tabular}} & \multirow{3}{*}{Saline-based} & \multirow{3}{*}{\SI{14}{bit} (\SI{510}{\nano\volt})} & \multirow{3}{*}{\textasciitilde\$1,500} & \multirow{3}{*}{\textasciitilde\SI{40}{min}} \\
        NeuroPath-DS-16C & 12 & 16 & & & & & \\
        NeuroPath-DS-8C  & 12 & 8  & & & & & \\
        \midrule

        BCIC-2a~\cite{brunner2008bci} & 9 & 22 & \multirow{2}{*}{\begin{tabular}{@{}c@{}}Medical-grade \\ (\eg g.tec)\end{tabular}} & \multirow{2}{*}{Gel-based Ag/AgCl} & \multirow{2}{*}{\begin{tabular}{@{}c@{}}High-precision \\ ($< \SI{1}{\micro\volt}$)\end{tabular}} & \multirow{2}{*}{$>\$\num{30000}$} & \textasciitilde\SI{58}{min} \\
        BCIC-2b~\cite{leeb2008bci} & 9 & 3  & & & & & \textasciitilde\SI{70}{min} \\
        \addlinespace 
        MI-KU~\cite{lee2019eeg}   & 19 & 56 & \begin{tabular}{@{}c@{}}Medical-grade \\ (BrainAmp)\end{tabular} & Gel-based Ag/AgCl & \SI{24}{bit} (\textasciitilde\SI{0.4}{\nano\volt}) & $>\$\num{50000}$ & \textasciitilde\SI{47}{min} \\
        
        \bottomrule
    \end{tabularx}
\end{table*}

\parahead{Data collection.} To comprehensively evaluate the performance of \systemname, we use six datasets, including three self-collected datasets and three benchmark public datasets as shown in Table~\ref{tab:datasets}.

\begin{figure}[t]
	\centering
	\includegraphics[width=2.8in]{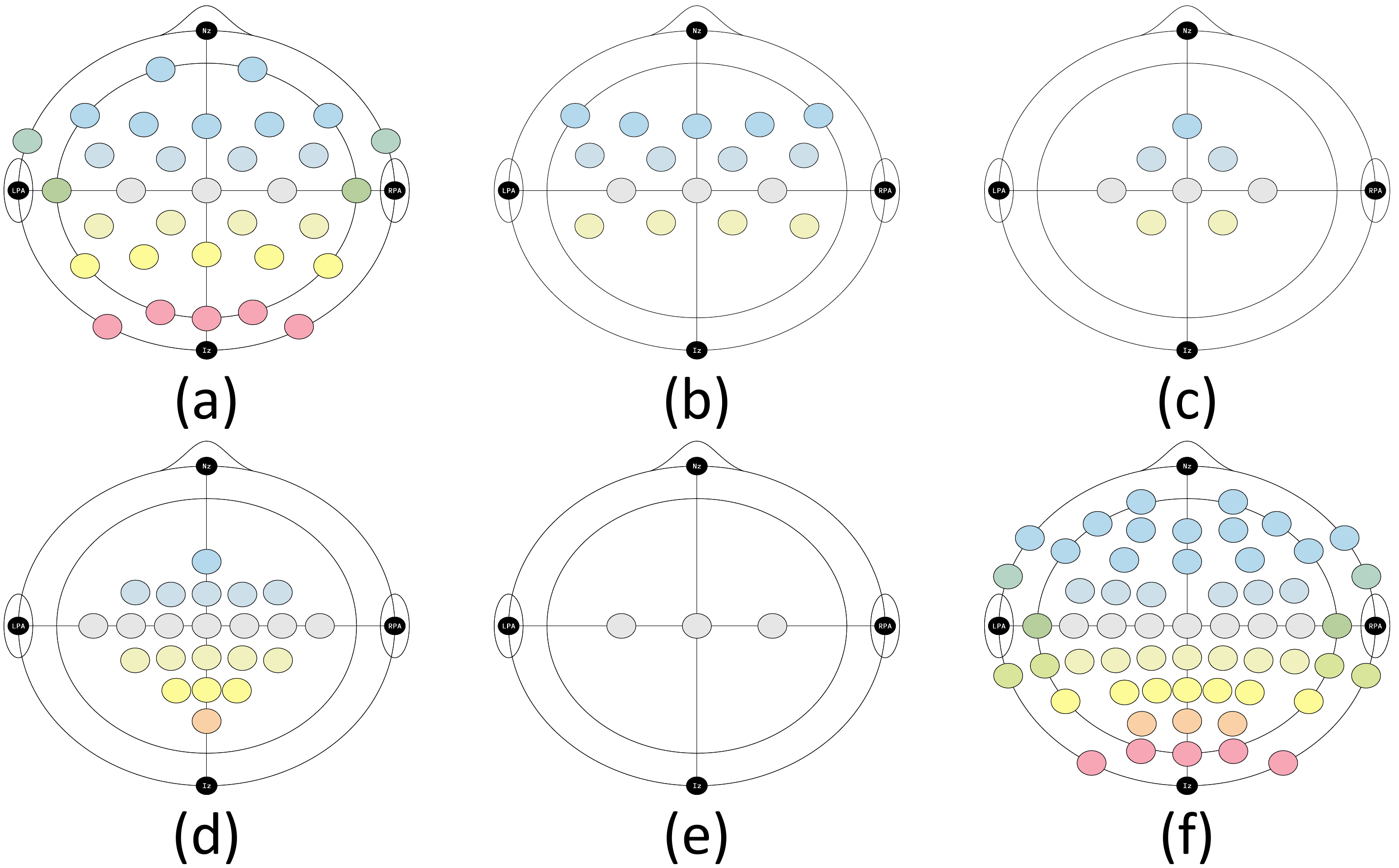}
	\vspace{-.1in}
        \caption{Different electrode numbers and placements for (a) NeuroPath-DS-32C, (b) NeuroPath-DS-16C, (c) NeuroPath-DS-8C, (d) BCIC-2a~\cite{brunner2008bci}, (e) BCIC-2b~\cite{leeb2008bci} and (f) MI-KU~\cite{lee2019eeg}.}
	\label{fig:placement}
	\vspace{-.1in}
\end{figure}

\textbf{1) NeuroPath-DS (Self-Collected).} This dataset is collected using the Emotiv FLEX 2~\cite{Emotiv}, a consumer-grade device. This device supports flexible channel configurations up to 32 channels and has a relatively coarse signal resolution (\ie 510 nV) compared to medical-grade equipment (as low as 0.39 nV). To systematically evaluate the performance of NeuroPath under varying levels of practicality and to comprehensively investigate the trade-off between channel count and decoding accuracy, we construct three sub-datasets with different numbers of channels. These sub-datasets represent full-channel, balanced, and high-portability application scenarios. Their respective electrode layouts are shown in Figure~\ref{fig:placement}(a), (b), and (c):

\begin{itemize}
    \item \textbf{NeuroPath-DS-32C:} A full-channel configuration that includes all 32 channels supported by the device. It is designed to serve as a performance benchmark and explore the system's potential.
    \item \textbf{NeuroPath-DS-16C:} A balanced configuration with 16 channels. These channels are selected based on their relevance to motor imagery, prioritizing those covering the sensorimotor cortex to balance decoding performance and ease of wear. 
    \item \textbf{NeuroPath-DS-8C:} A high-portability configuration with only 8 channels. The channel selection focuses on the most critical brain regions for MI (\eg near C3, C4, and Cz), representing a quick-setup, daily-use consumer scenario.
\end{itemize} 

\begin{figure}[b]
	\centering
	\includegraphics[width=3in]{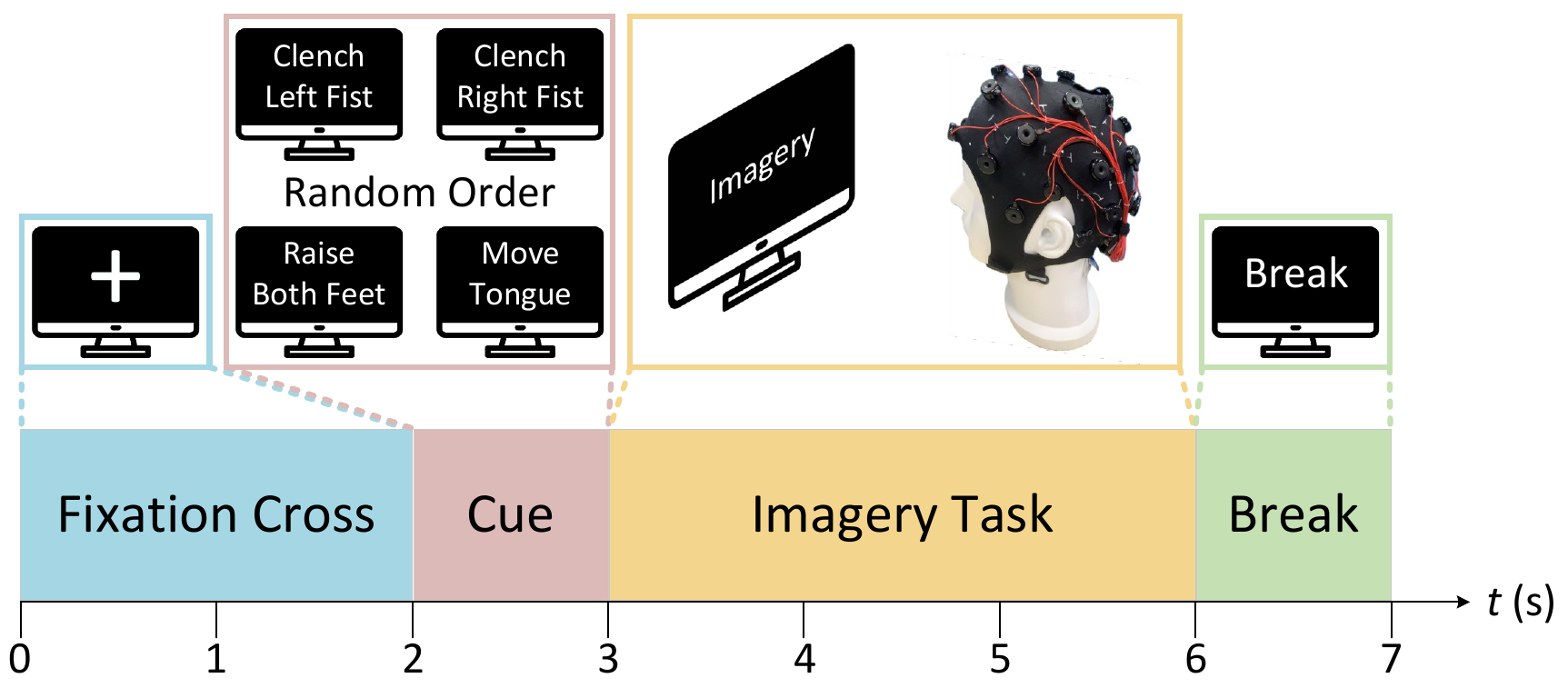}
	\vspace{-.1in}
        \caption{Data collection procedure of one trial in self-collected NeuroPath-DS. This process is similar to that of public medical-grade datasets, such as BCIC-2a~\cite{brunner2008bci}.}
	\label{fig:exp_setting}
\end{figure}

Data for all sub-datasets are collected from twelve volunteers (four females and eight males), with ethical approval obtained from our institution. Before formal data collection, each volunteer performs the actual movements to activate the relevant motor cortex regions and to build an intuitive understanding of the imagery tasks. During data collection, users are explicitly instructed to keep their body physically still and perform mental rehearsal of the action. The procedure for one trial is shown in Figure~\ref{fig:exp_setting}. Volunteers sit comfortably facing a computer screen. At $t$=0 s, a fixation cross appears. At $t$=2 s, a prompt (\eg ``clench left fist'', ``clench right fist'', ``raise both feet'', or ``move tongue'') is displayed, and the order of prompts is randomized across trials. This visual instructional cue serves as the ground truth for the user's motor intention in each trial. At $t$=3 s, the prompt disappears and the volunteer begins the corresponding motor imagery, which continues until $t$=6 s, when a ``break'' message indicates the trial's end. We use the 4-second EEG segment from $t$=2 s to $t$=6 s as the effective imagery data for training and testing. The data are then split into training and testing sets at a 4:1 ratio to train a personalized model for each volunteer.

\textbf{2) BCIC-2a~\cite{brunner2008bci}.} This public dataset contains EEG data from 9 subjects, recorded using 22 Ag/AgCl electrodes (a type of high-quality wet electrode) with the placement shown in Figure~\ref{fig:placement}(d) at a sampling rate of 250 Hz. The paradigm is a four-class MI task involving the imagined movement of the left hand, right hand, both feet, and tongue. During a trial, a fixation cross appears at $t$=0 s. At $t$=2 s, a visual cue lasting 1.25 s instructs the subject to perform the imagery task, which continues until $t$=6 s. The cues for each class appear in a random order. We use the 4-second data from $t$=2 s to $t$=6 s as the effective imagery data. Each subject's data includes a training and an evaluation session, each with 288 trials.

\textbf{3) BCIC-2b~\cite{leeb2008bci}.} This public dataset is also recorded from 9 subjects, using 3 Ag/AgCl electrodes (same high-quality type as in BCIC-2a) with the placement shown in Figure~\ref{fig:placement}(e) at a 250 Hz sampling rate. The paradigm is a two-class MI task involving imagined movement of the left and right hand. The procedure is similar, with subjects performing a 4-second MI task after a visual cue. The cues for each class appear in a random order. Each subject's data consists of five sessions, with each session containing 120 to 160 trials, split equally between the two classes. Data from the first three sessions constitute the training set, while data from the final two sessions form the evaluation set.

\textbf{4) MI-KU~\cite{lee2019eeg}.} This public dataset is collected using a BrainAmp amplifier, a device offering high-precision 24-bit resolution and excellent noise rejection, with 62 Ag/AgCl electrodes arranged as shown in Figure~\ref{fig:placement}(f), at a sampling rate of 1000 Hz. As only 56 of its electrodes conform to the international 10-10 standard system layout, we use only these 56 channels in our experiments. The dataset includes 54 subjects. The original paper notes that some subjects exhibit ``MI BCI illiteracy'', meaning they cannot effectively produce distinguishable MI signals. To ensure the quality of our pre-training data and align with the paper's findings, we select data from 19 subjects who are not ``MI BCI illiterate'' for our experiments. The paradigm is a two-class MI task involving the imagined movement of the left hand and right hand. Each trial begins with a 3-second preparation period, followed by a visual cue and a 4-second imagery task. The cues for each class appear in a random order. Each subject's data includes two sessions, each containing a training phase and a testing phase, with 100 trials per phase (50 for each of the two classes).

\parahead{Evaluation metric.} We use accuracy as the evaluation metric. Accuracy is defined as the percentage of test trials correctly decoded out of the total number of test trials.

\subsection{Overall Performance}
\label{exp:op}

We compare the following methods:

\begin{itemize}
    \item \textbf{LGL-BCI~\cite{lu2025lgl}.} This is a recent end-to-end geometric learning approach designed for MI decoding.
    \item \textbf{CLTNet~\cite{gu2025cltnet}.} This is a recent end-to-end hybrid deep learning architecture designed for MI decoding.
    \item \textbf{CIACNet~\cite{liao2025composite}.} This is a recent end-to-end composite improved attention convolutional network designed for MI decoding.
    \item \textbf{\systemname.} The system proposed in this paper.
\end{itemize}

\begin{figure}[t]
	\centering
	\includegraphics[width=3.2in]{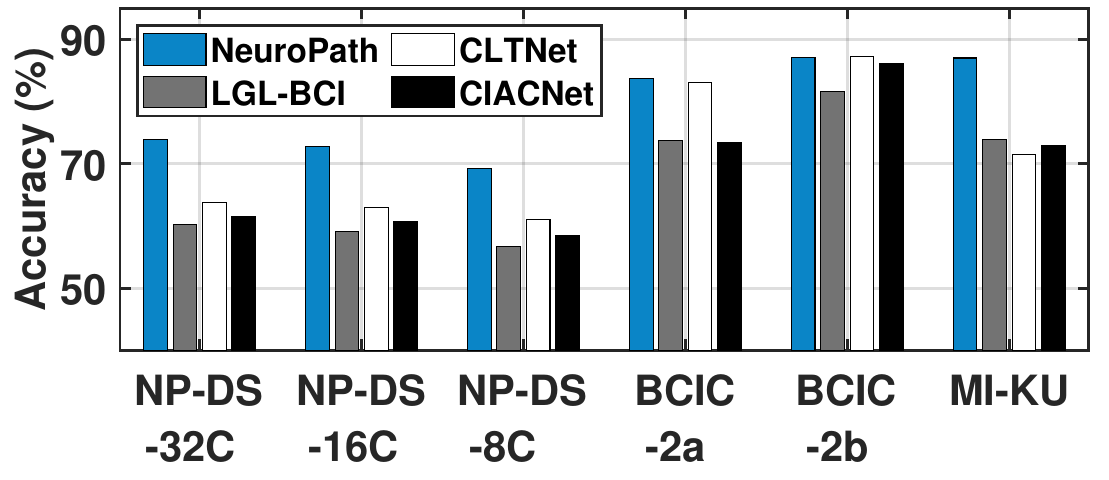}
	\vspace{-.1in}
        \caption{Overall performance. ``NP'' means ``\systemname''.}
	\label{fig:overall}
	\vspace{-.15in}
\end{figure}

Figure~\ref{fig:overall} shows the overall performance of the four methods across six  different datasets. Specifically, we first pre-train the shared MI decoder on the three medical-grade public datasets BCIC-2a, BCIC-2b, and MI-KU using a cross-dataset pre-training framework, which enables it to learn rich prior knowledge. Subsequently, we fine-tune the pre-trained model on each of the six target datasets. During the fine-tuning phase, we simultaneously activate the cross-modality knowledge transfer framework to further enhance the model's decoding performance on specific tasks. Throughout the entire process, we strictly ensure that there is no overlap between the training and test sets.

We can see that \systemname consistently maintains competitive performance compared to the other three baselines on all six datasets with different electrode numbers and placements. To be specific, on the three challenging consumer-grade datasets (\systemname-DS with 32, 16, and 8 channels), \systemname's decoding accuracies are 73.9\%, 72.8\%, and 69.2\%, respectively. These results significantly outperform LGL-BCI by margins of 12.5\%--13.7\%, CLTNet by 8.1\%--10.2\%, and CIACNet by 10.8\%--12.4\%. Furthermore, on the three medical-grade public datasets (BCIC-2a, BCIC-2b, and MI-KU), \systemname achieves decoding accuracies of 83.7\%, 87.1\%, and 87.0\%, respectively, which outperform or are highly comparable to LGL-BCI (by 5.5\%--13.1\%), CLTNet (by -0.2\%--15.5\%), and CIACNet (by 1.0\%--14.0\%). Ultimately, unlike previous end-to-end methods, \systemname aims not only to achieve high decoding performance but also to address the critical challenges of practical deployment in heterogeneous electrode configurations and low-SNR scenarios.

\subsection{Ablation Study.}
\label{exp:as}

We conduct an ablation study to systematically evaluate the effectiveness of each module in \systemname. To this end, we construct and evaluate the following intermediate versions of \systemname on all six datasets:

\begin{itemize}
    \item \textbf{V1:} This version serves as the baseline, training the MI decoder (\S\ref{design:decoder}) end-to-end on each dataset independently.

    \item \textbf{V2-1:} Building on V1, this version applies the cross-dataset pre-training framework from \S\ref{design:enhancement:same}. We first pre-train the shared MI decoder on the three medical-grade public datasets (BCIC-2a, BCIC-2b, and MI-KU) and then fine-tune it on each of the six target datasets.

    \item \textbf{V2-2:} Building on V1, this version introduces the cross-modality knowledge transfer framework from \S\ref{design:enhancement:different}, which leverages generated skeleton data as a teacher signal to enhance the MI decoder's performance.

    \item \textbf{V3 (Full):} This is the complete version of \systemname, integrating all proposed modules.
\end{itemize}

\begin{table}[h]
\centering
\caption{Ablation study.}
\vspace{-.1in}
\label{tab:ablation}
\begin{tabular}{ccccc}
\toprule
\textbf{Dataset} & \textbf{V1} & \textbf{V2-1} & \textbf{V2-2} & \textbf{V3 (Full)}\\
\midrule
\textbf{NeuroPath-DS-32C} & 67.0\% & 71.6\% & 69.3\% & \textbf{73.9\%}\\
\textbf{NeuroPath-DS-16C} & 66.8\% & 69.3\% & 68.7\% & \textbf{72.8\%}\\
\textbf{NeuroPath-DS-8C} & 66.1\% & 68.6\% & 68.5\% & \textbf{69.2\%}\\
\textbf{BCIC-2a} & 78.1\% & 82.4\% & 80.1\% & \textbf{83.7\%}\\
\textbf{BCIC-2b} & 85.8\% & 86.3\% & 85.9\% & \textbf{87.1\%}\\
\textbf{MI-KU} & 78.4\% & 85.9\% & 81.3\% & \textbf{87.0\%}\\
\bottomrule
\end{tabular}
\vspace{-.1in}
\end{table}

As shown in Table~\ref{tab:ablation}, the baseline version (V1) achieves decoding accuracies of 67.0\%, 66.8\%, and 66.1\% on the three consumer-grade NeuroPath-DS datasets, and 78.1\%, 85.8\%, and 78.4\% on the three public datasets, respectively. With V2-1, the accuracies increase to 71.6\%, 69.3\%, and 68.6\% on the NeuroPath-DS datasets, and to 82.4\%, 86.3\%, and 85.9\% on the public datasets. This demonstrates that by pre-training on larger and higher-quality datasets, the model learns more robust neural feature priors. This knowledge is then effectively transferred to the target task during the fine-tuning phase, thereby enhancing decoding performance. Similarly, solely applying the cross-modality knowledge transfer strategy (V2-2) also yields comprehensive performance gains, boosting the accuracies to 69.3\%, 68.7\%, and 68.5\% on the NeuroPath-DS datasets, and to 80.1\%, 85.9\%, and 81.3\% on the public datasets. This proves that transferring knowledge from the clear and feature-rich skeleton modality can effectively guide the MI decoder in learning more discriminative motor representations from noisy EEG signals. Finally, the complete version of \systemname (V3), which integrates all modules, achieves the best performance across all datasets, reaching accuracies of 73.9\%, 72.8\%, and 69.2\% on the NeuroPath-DS datasets, and 83.7\%, 87.1\%, and 87.0\% on the public datasets. This indicates a positive synergy between the two enhancement strategies, as their combination maximizes the model's decoding capability.

\subsection{Micro-benchmarks}
\label{exp:micro}

We then conduct micro-benchmark experiments to evaluate how the performance of \systemname is impacted by various factors.

\begin{figure*}[t]
\begin{minipage}[t]{0.32\linewidth}
\centering
\includegraphics[width=1.0\linewidth]{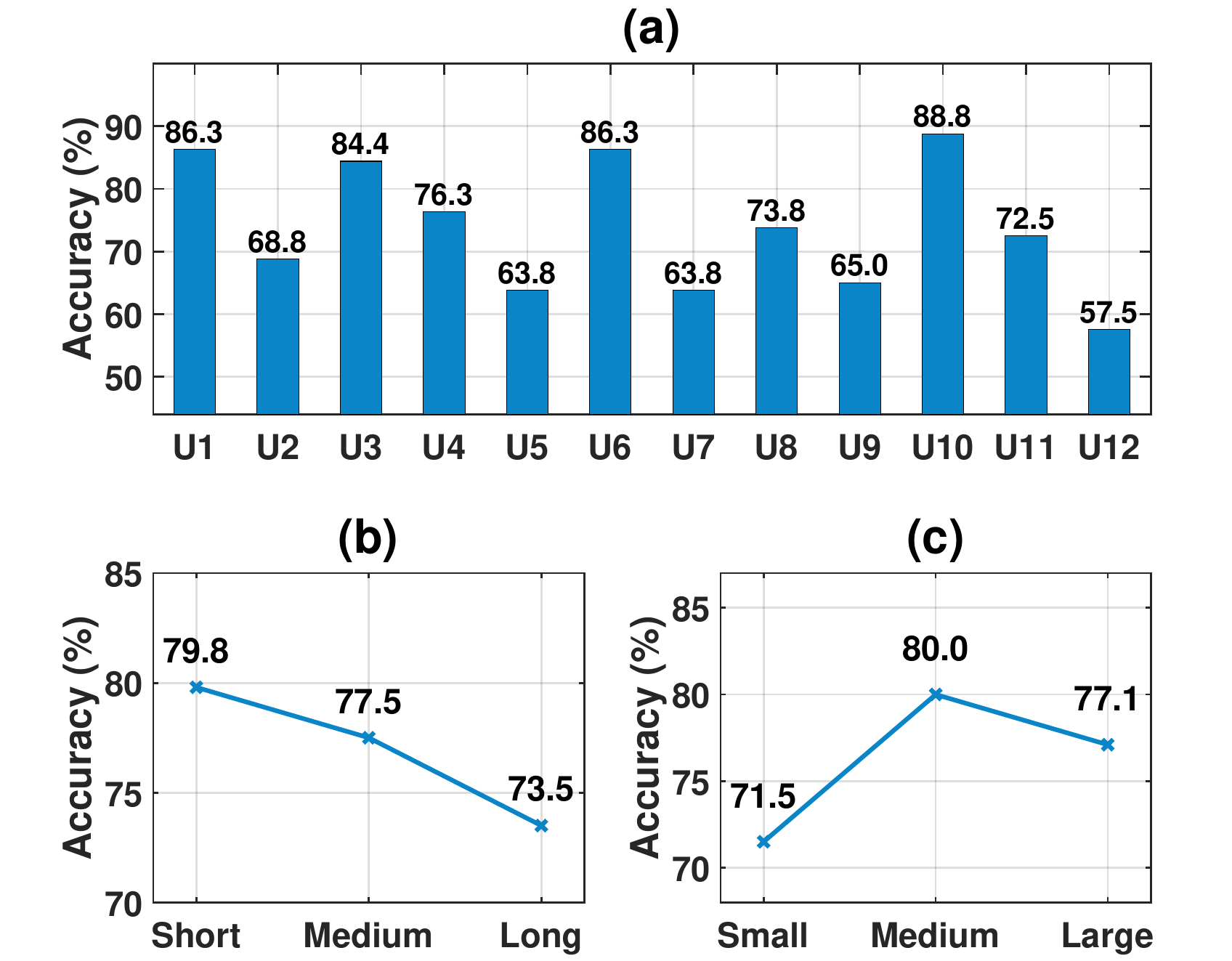}
\vspace{-.3in}
\caption{Performance across different (a) users, (b) hair lengths, and (c) head circumferences.}
\label{fig:user}
\end{minipage}
\hspace{0.1cm}
\begin{minipage}[t]{0.32\linewidth}
\centering
\includegraphics[width=1.0\linewidth]{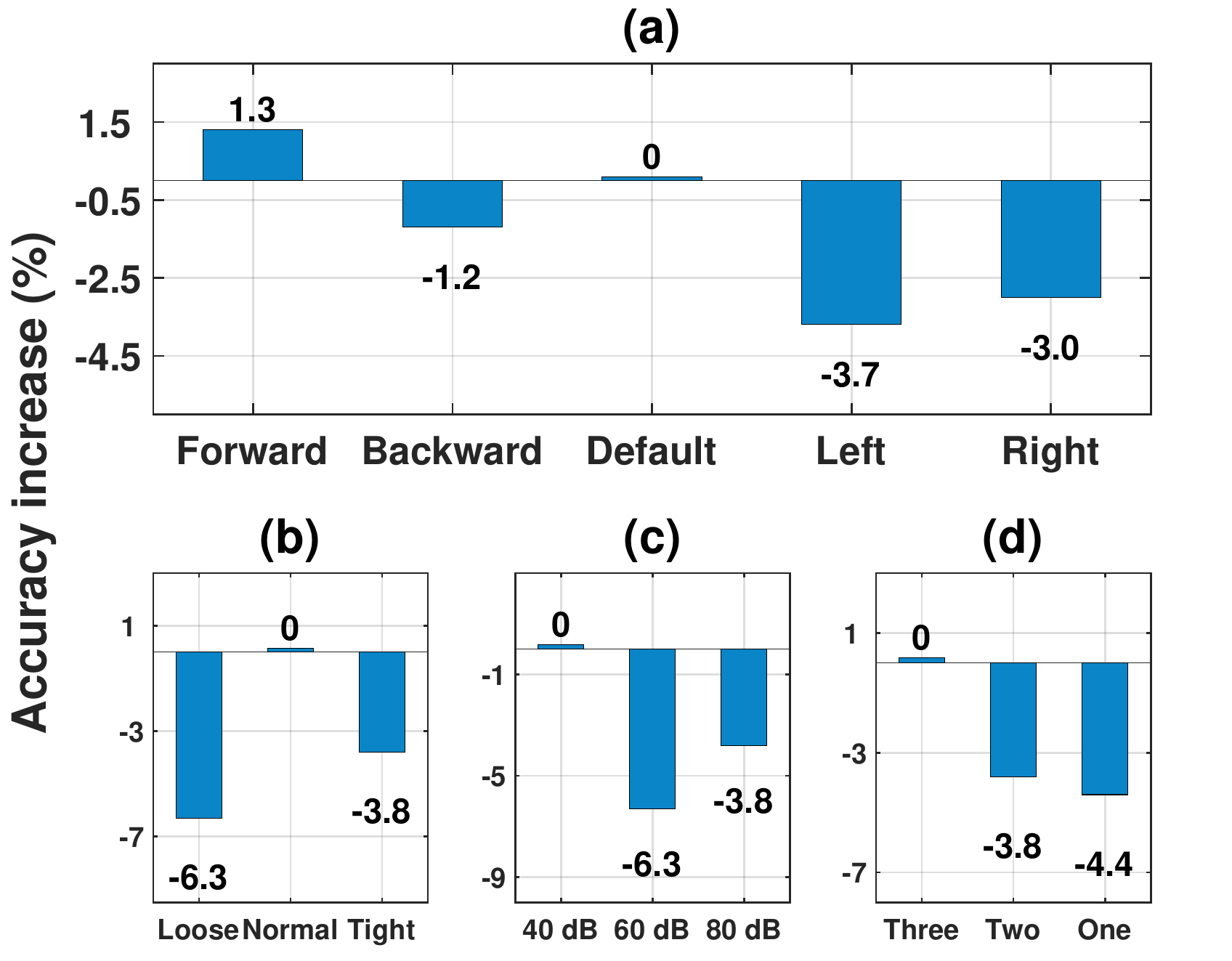}
\vspace{-.3in}
\caption{Performance under varying (a) wearing positions, (b) sensor attachments, (c) ambient volumes, and (d) pre-training data scales.}
\label{fig:position}
\end{minipage}
\hspace{0.1cm}
\begin{minipage}[t]{0.32\linewidth}
\centering
\includegraphics[width=1.0\linewidth]{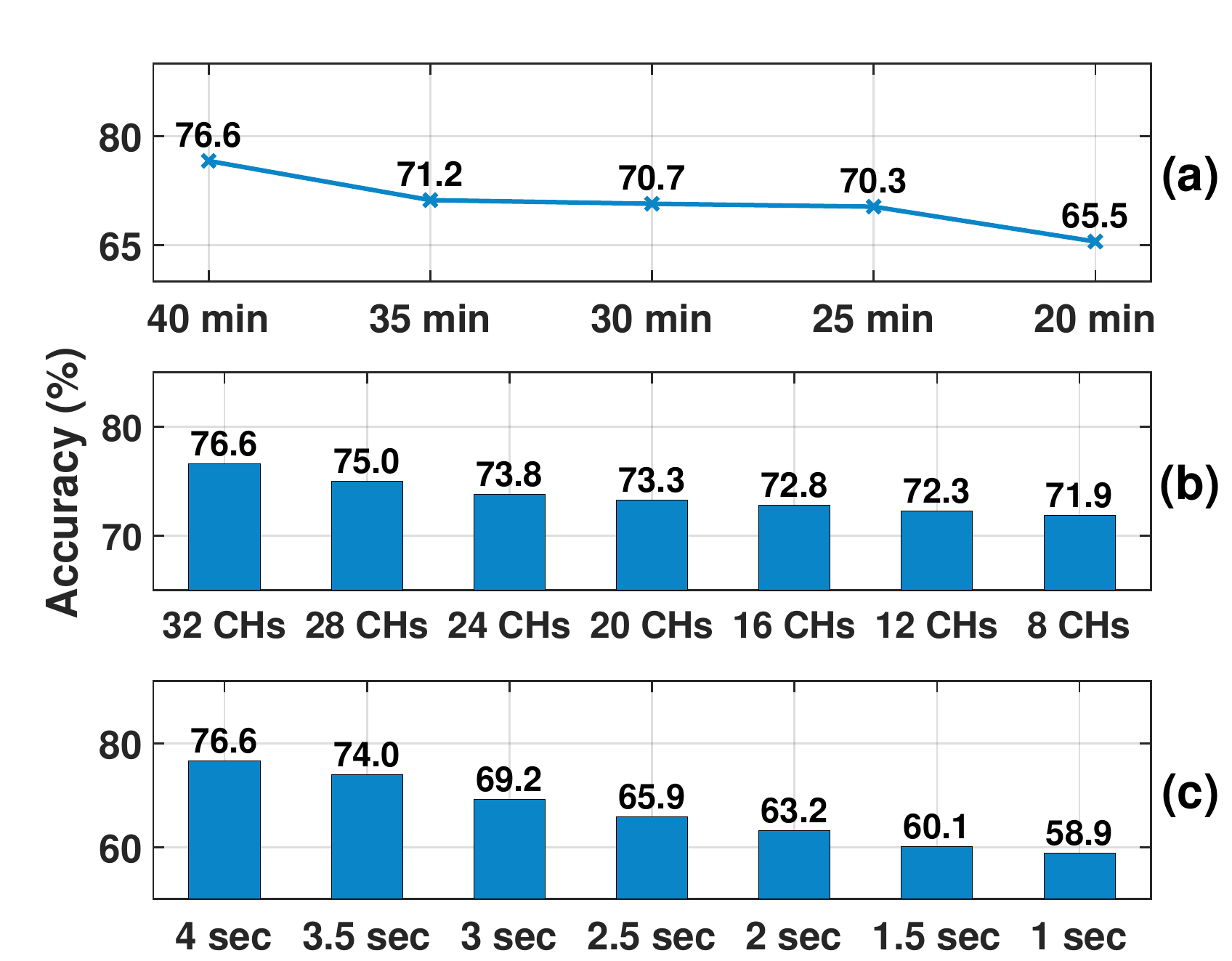}
\vspace{-.3in}
\caption{Performance with different (a) amounts of training data, (b) numbers of electrodes, and (c) trial data lengths.}
\label{fig:amounts}
\end{minipage}
\vspace{-.15in}
\end{figure*}

\parahead{Different users.} Figure~\ref{fig:user}(a) shows the performance of \systemname on the twelve different users in the NeuroPath-DS-32C dataset. It is evident that there are individual differences in decoding accuracy, ranging from 57.5\% to 88.8\%. This variance primarily stems from inherent differences in physiological structures and neural activity patterns among users, which is a common phenomenon in the BCI field~\cite{lee2019eeg}. This performance can be further enhanced by collecting more user-specific data, which is a practical approach for \systemname, as ground truth acquisition is straightforward and only requires the user to wear the device while following instructions.

\parahead{Different hair lengths.} We categorize users into three groups based on hair length: Short (< 5 cm), Medium (5-30 cm), and Long (> 30 cm). Figure~\ref{fig:user}(b) shows the average decoding performance for each group. The short-haired group achieves the highest accuracy at 79.8\%, while the accuracies for the medium and long-haired groups decrease to 77.5\% and 73.5\%, respectively. This is because longer hair may form an insulating layer between electrodes and the scalp, which increases contact impedance and thereby degrades the quality of the EEG signal, affecting decoding performance.

\parahead{Different head circumferences.} We divide users into three groups based on head circumference: Small (male <56 cm, female <54 cm), Medium (male: [56 cm, 58 cm], female: [54 cm, 56 cm]), and Large (male >58 cm, female >56 cm). Figure~\ref{fig:user}(c) shows the average performance for each group. The medium circumference group performs best, reaching 80.0\%, while the performance for the small and large circumference groups drops to 71.5\% and 77.1\%, respectively. This is likely because the fixed-size EEG cap we use provides the best fit for users with medium head circumferences. A head that is too small may result in poor electrode contact, while a head that is too large may cause discomfort from the overly tight cap, affecting signal quality. This issue can be addressed by providing a custom-fitted cap for each user.

\parahead{Different wearing positions.} In our experiments, we use the Cz electrode (at the center of the scalp) as the anchor for the default wearing position. We then shift the entire cap forward, backward, left, and right by 2 cm to simulate slight positional deviations that might occur in daily use. As shown in Figure~\ref{fig:position}(a), compared to the default position, performance slightly improves when the cap is worn backward but decreases when it is moved in other directions, with a leftward shift having the largest negative impact (-3.7\%). Overall, the impact of minor positional shifts on performance is within an acceptable range, indicating that our design, particularly the graph attention adapter, can dynamically focus on important electrode information and has a degree of robustness to slight changes in position.

\parahead{Variations in strap tightness.} We designate the state where users adjust the EEG cap to a comfortable level as ``Normal''. Building on this, we loosen and tighten the straps by 1 cm, labeling these states as ``Loose'' and ``Tight'', respectively. As shown in Figure~\ref{fig:position}(b), performance drops in the ``Loose'' state (-6.3\%), as the poor fit leads to inadequate contact between the electrodes and the scalp, degrading signal quality. Performance also decreases in the ``Tight'' state (-3.8\%), possibly because the excessive tightness causes user discomfort or muscle tension, which introduces additional noise.

\parahead{Different ambient volumes.} We evaluate the performance in different noise environments by simulating three scenarios: 40 dB (a quiet private bedroom), 60 dB (an office with people talking), and 80 dB (extreme weather with thunder outside). As shown in Figure~\ref{fig:position}(c), performance is best in the quiet environment (40 dB). As the ambient volume increases, performance declines, with accuracy decreasing by 6.3\% and 3.8\% in the 60 dB and 80 dB environments, respectively. This suggests that external noise, especially distracting conversations, can interfere with the user's ability to concentrate on motor imagery, thereby affecting decoding performance.

\parahead{Different amounts of pre-training data.} We evaluate the impact of pre-training data scale by training on one, two, or all three public datasets. Compared to the three-dataset baseline, average decoding accuracy decreases by 3.8\% and 4.4\% when reducing the pre-training scale to two and one dataset, respectively (Figure~\ref{fig:position}(d)). This demonstrates that a larger and more diverse pre-training volume enables the shared decoder to learn more generalized and robust neural feature priors, significantly benefiting downstream performance.

\parahead{Different amounts of training data.}
Since collecting EEG data is time-consuming and demanding for users, an ideal MI system must be data-efficient. To evaluate this, we examine the sensitivity of \systemname to the amount of training data by training the model with 40, 35, 30, 25, and 20 minutes of data per user. As shown in Figure~\ref{fig:amounts}(a), performance decreases as the training data are reduced. Notably, reducing the data from 40 minutes to 25 minutes (a 37.5\% reduction) only lowers accuracy from 76.6\% to 70.3\%. Even with just 20 minutes of training data (a 50\% reduction), the accuracy remains 65.5\%. These results demonstrate that \systemname maintains good performance under limited data. Moreover, it suggest that with more training data, the model has the potential to further improve its performance.

\parahead{Different numbers of electrodes.} The number of electrodes on a wearable EEG cap represents a critical trade-off between convenience and performance. We start with the default 32 electrodes and progressively reduce the number down to 8, prioritizing the retention of electrodes near the motor cortex area in each step. Figure~\ref{fig:amounts}(b) shows the decoding accuracy decreases gracefully as the number of electrodes is reduced. When the electrode count is lowered from 32 to 8, the accuracy drops from 76.6\% to 71.9\%, a performance loss that is within an acceptable range. This indicates that while a dense electrode layout provides richer information, \systemname can also operate effectively with fewer electrodes. This provides support for the future design of more portable, consumer-grade devices with fewer electrodes.

\parahead{Different data lengths.} In this experiment, we comprehensively evaluate the performance of \systemname with varying data lengths by reducing the trial length from the standard 4 seconds down to 1 second. As shown in Figure~\ref{fig:amounts}(c), when the trial length is reduced from 4 seconds to 1 second, the accuracy drops from 76.6\% to 58.9\%. This result indicates that \systemname maintains effective decoding capabilities when operating on certain shorter data segments (\eg greater than 2 seconds). However, when the window is further reduced (below 2 seconds), the accuracy experiences a more noticeable decline. This objectively illustrates the inherent trade-off between low latency and high decoding accuracy.

\parahead{BCI illiterate population.} In this experiment, we evaluate the performance of the remaining users in the MI-KU dataset, specifically the user group considered to be BCI illiterate. BCI illiteracy is a recognized physiological phenomenon in the field, referring to users who inherently cannot produce motor imagery EEG patterns that can be effectively detected by the system. We follow the general consensus in the field, specifically applying a decoding accuracy threshold below 75\% to the subjects in the MI-KU dataset to define its BCI illiterate population. As shown in Table~\ref{tab:bci_illiteracy_detailed}, the performance of this population (35 subjects in total) ranges from 54.0\% to 74.0\%, with an average accuracy of 62.2\%. This physiological limitation poses a concrete impact on real-world deployment, highlighting the boundary of current MI-BCI applicability.

\begin{table}[htbp]
\centering
\caption{Performance on the BCI illiterate population.}
\vspace{-.1in}
\label{tab:bci_illiteracy_detailed}
\resizebox{\columnwidth}{!}{
\begin{tabular}{lc|lc|lc|lc|lc}
\toprule
\textbf{ID} & \textbf{Acc(\%)} & \textbf{ID} & \textbf{Acc(\%)} & \textbf{ID} & \textbf{Acc(\%)} & \textbf{ID} & \textbf{Acc(\%)} & \textbf{ID} & \textbf{Acc(\%)} \\
\midrule
S01 & 62.5 & S05 & 64.0 & S07 & 65.5 & S08 & 65.0 & S09 & 57.5 \\
S10 & 57.0 & S11 & 66.0 & S12 & 68.0 & S13 & 61.5 & S14 & 56.5 \\
S15 & 66.5 & S17 & 54.5 & S19 & 63.0 & S20 & 71.5 & S22 & 57.5 \\
S23 & 57.0 & S24 & 60.5 & S28 & 63.0 & S31 & 59.5 & S32 & 71.0 \\
S34 & 54.0 & S37 & 73.0 & S38 & 59.0 & S39 & 55.0 & S40 & 64.0 \\
S41 & 59.0 & S45 & 74.0 & S46 & 74.0 & S48 & 63.0 & S49 & 55.5 \\
S50 & 54.0 & S51 & 57.5 & S52 & 68.5 & S53 & 57.0 & S54 & 61.0 \\
\bottomrule
\end{tabular}
}
\end{table}

\subsection{System Overhead} To evaluate real-world practicality, we deploy \systemname on a Google Pixel 7 (Android 13, 4355 mAh battery) via PyTorch Mobile. Table~\ref{tab:overhead} summarizes the system overhead metrics.

\begin{table}[h]
\vspace{-.1in}
\centering
\caption{System overhead of \systemname.}
\vspace{-.1in}
\label{tab:overhead}
\begin{tabular}{@{}ll@{}}
\toprule
\textbf{Metric}                  & \textbf{Value}                 \\ \midrule
Model Parameters                 & 60,176                         \\
Computational Complexity         & 352.17 M-Adds                  \\
Inference Latency (per trial)    & 21.22 ms                       \\
Power Consumption                & 145.10 mAh                      \\
Battery Impact                   & 3.35\% of total battery        \\ \bottomrule
\end{tabular}
\vspace{-.1in}
\end{table}

\textbf{1) Computational cost.} The model in \systemname is exceptionally lightweight, requiring only 60.2 K parameters and 352.17 M-Adds per inference. This load is easily manageable by standard mobile CPUs without specialized hardware accelerators. \textbf{2) Real-time performance.} \systemname processes a complete 4-second MI trial in just 21.22 ms (averaged over 100 runs). This latency constitutes less than 0.6\% of the data window's duration (21.22 ms / 4000 ms), ensuring that the system can provide immediate feedback without any perceptible delay to the user. \textbf{3) Energy efficiency.} Measured via Android's Batterystats and Battery Historian~\cite{Battery}, the power consumption is 145.10 mAh (3.35\% of total capacity). This low footprint supports nearly 30 hours of continuous operation on a single charge, confirming its viability for daily use.

%% file: Body/related.tex
\section{Discussion}
\label{sec:Discussion}

\parahead{Physiological limitations and BCI illiteracy.} A fundamental challenge for general deployment is BCI illiteracy, which refers to the inherent inability of certain users to produce distinct and detectable EEG patterns during motor imagery. Currently, \systemname cannot resolve this physiological limitation. Future solutions may require significantly extending user BCI training to stimulate neural responses, or integrating alternative control channels (\eg eye-tracking or electromyography) to ensure universal accessibility.

\parahead{Robustness against physical motion.} A current limitation of \systemname is its inability to handle strong physical motion, as muscle-generated electrical signals easily overpower subtle MI neural oscillations. However, since the MI paradigm primarily targets individuals with severe motor impairments, scenarios involving active physical movement fall largely outside its intended scope.

\parahead{Scalability to complex control.} The number of classes in current evaluation is relatively small. However, this setup is consistent with mainstream protocols in the motor imagery field and largely addresses the core interaction needs of typical BCI applications, such as wheelchair control and discrete smart home interactions. In the future, with hardware upgrades and improved signal resolution, it is possible to extend this framework to more complex, fine-grained, or continuous control scenarios.

\section{Related Work}
\label{sec:related}
We review the related work of this paper in this section.

\parahead{Brain-Computer Interface (BCI) paradigms.}
Non-invasive EEG-based BCIs are highly portable and suitable for consumer use~\cite{defossez2023decoding,ouyang2024neurobci,wu2022deepbrain}. While stimulus-driven paradigms like SSVEP~\cite{guo2022ssvep} and P300~\cite{herrmann2001mechanisms} cause user fatigue~\cite{azadi2023fatigue}, Motor Imagery (MI) provides an intuitive, internally-driven control channel~\cite{marchesotti2016quantifying,schuster2011best}. However, widespread MI-BCI deployment is hindered by complex signal generation and the inherently low SNR of consumer-grade EEG~\cite{bear2020neuroscience}.

\parahead{Motor Imagery (MI) decoding methods.}
Early MI decoding relied on feature engineering (\eg CSP~\cite{jiang2020efficient}), while recent deep learning models directly learn spatiotemporal features~\cite{schirrmeister2017deep,lawhern2018eegnet,altaheri2022physics}. Despite high accuracy, existing models face critical limitations: (1) they lack neurophysiological interpretability~\cite{lu2025lgl,zhang2018converting}; (2) they are typically trained on medical-grade datasets and fail to generalize to the low-quality signals of consumer devices~\cite{zhao2024ctnet}; and (3) they assume fixed electrode layouts~\cite{altaheri2023deep,gu2025cltnet,liao2025composite}. To address these gaps, \systemname proposes a brain-inspired architecture paired with cross-modality and cross-dataset training strategies, ensuring robustness across heterogeneous, low-quality datasets.

\parahead{Practicality of mobile sensing systems.}
Enhancing consumer-grade hardware practicality is central to mobile sensing~\cite{chen2024exploring,ji2023construct,yang2025hearforce,cao2024practical}. Moreover, the broader ecosystem of these mobile systems is continuously enriched by system-level advances in edge inference~\cite{wang2024swapnet,shen2024fedconv,wang2024latte,shen2025gpiot} and scalable low-power wireless communications~\cite{hou2025molora,yu2024revolutionizing,hou2024one}. Prior work extracts robust features from weak signals using COTS devices~\cite{cao2024finger,hu2024breathpro,butkow2024evaluation}, advanced signal processing~\cite{li2025ceiverfi,yun2024powdew,liu2025respear}, physical modeling~\cite{mehrotra2024hydra,cao2022gaze}, and robust learning strategies~\cite{ouyang2024admarker,nie2025soundtrack,ji2022sifall,liu2025earmeter}. Inspired by these works, we tackle the unsolved challenge of robust decoding on low-quality consumer EEG by utilizing neurophysiology as a high-level architectural inspiration and by developing training strategies that generalize across heterogeneous datasets in the design of \systemname in this paper.

%% file: Body/conclusion.tex
\section{Conclusion}
\label{sec:concl}

In this paper, we present \systemname, a well-designed neural architecture addressing practical Motor Imagery (MI) BCI challenges. Departing from conventional fragmented models, it introduces a unified deep learning framework inspired by the brain's signal generation pathway. A spatially aware graph adapter handles variable sensor configurations, while a multimodal auxiliary training strategy enhances robustness against low-SNR consumer-grade devices. Extensive evaluations across three public medical-grade and three self-collected consumer-grade datasets demonstrate that \systemname consistently outperforms recent methods.

\begin{acks}
Zhenjiang Li was supported by the GRF grants from Research Grants Council of Hong Kong (CityU 11205624 and CityU 11213622). Corresponding authors: Yang Liu and Zhenjiang Li.
\end{acks}